\documentclass[preprintnumbers,aps,twocolumn,prx,amsmath,amssymb,10pt,aps]{revtex4-2} 

\usepackage{graphicx}
\usepackage{latexsym}
\usepackage{amsmath}
\usepackage{wasysym}
\usepackage{amsthm}
\usepackage{amsbsy}
\usepackage{amssymb}
\usepackage{epstopdf}
\usepackage{enumerate}
\usepackage{setspace}
\usepackage{dcolumn}
\usepackage{bm}
\usepackage{slashed}
\usepackage{color}
\usepackage{youngtab}
\usepackage{multirow}
\usepackage{array}
\usepackage{mathtools}
\usepackage{inputenc}
\usepackage{makecell}
\usepackage{algpseudocode}
\usepackage{tikz}
\usetikzlibrary{quantikz2}

\usepackage[colorlinks=true,linkcolor=blue,citecolor=blue,urlcolor=blue]{hyperref}

\newcommand{\lvertline}[1]{\multicolumn{1}{|c}{#1}}
\newcommand{\rvertline}[1]{\multicolumn{1}{c|}{#1}}
\newcommand{\lrvertline}[1]{\multicolumn{1}{|c|}{#1}}

\begin{document}

\title{Scalable Quantum Algorithms for Gutzwiller Projection}

\author{Byungmin Kang$^{1,2}$, Hyunwoong Kwon$^{2}$, Vito W. Scarola$^{3}$ and Kwon Park$^{2}$}
\affiliation{$^{1}$Center for Theoretical Physics---a Leinweber Institute, Department of Physics, Massachusetts Institute of Technology, Cambridge, MA 02139, USA}
\affiliation{$^{2}$School of Physics, Korea Institute for Advanced Study, Seoul 02455, Korea}
\affiliation{$^{3}$Department of Physics, Virginia Tech, Blacksburg, VA 24061, USA}

\preprint{MIT-CTP/6038}
\date{\today}

\begin{abstract}
Quantum simulation requires highly accurate input states.  
Gutzwiller-projected Bardeen-Cooper-Schrieffer (BCS) states provide physically motivated input states for solving strongly correlated lattice models, but their preparation on a quantum computer is hindered by the non-trivial nature of the Gutzwiller projection. We construct scalable quantum algorithms for this task by combining a circuit construction for arbitrary BCS states with the amplitude amplification for Gutzwiller projection (AAGP) procedure. AAGP yields a quadratic reduction in the number of projection queries compared with measurement-based postselection and leads to substantially improved fault-tolerant resource scaling. For projected BCS states optimized for the square-lattice $t$-$J$ model, we find that the projected-state weight decreases exponentially with system size, but the quadratic improvement is still large enough at physically relevant finite sizes to make a decisive practical difference. In particular, for a 100-site benchmark, AAGP reduces the required number of projection queries by about seven orders of magnitude. These results establish AAGP as an enabling input-state preparation protocol for projected BCS states in quantum simulation. 
\end{abstract}

\maketitle

\begin{figure}[th!]
\includegraphics[width=0.44\textwidth]{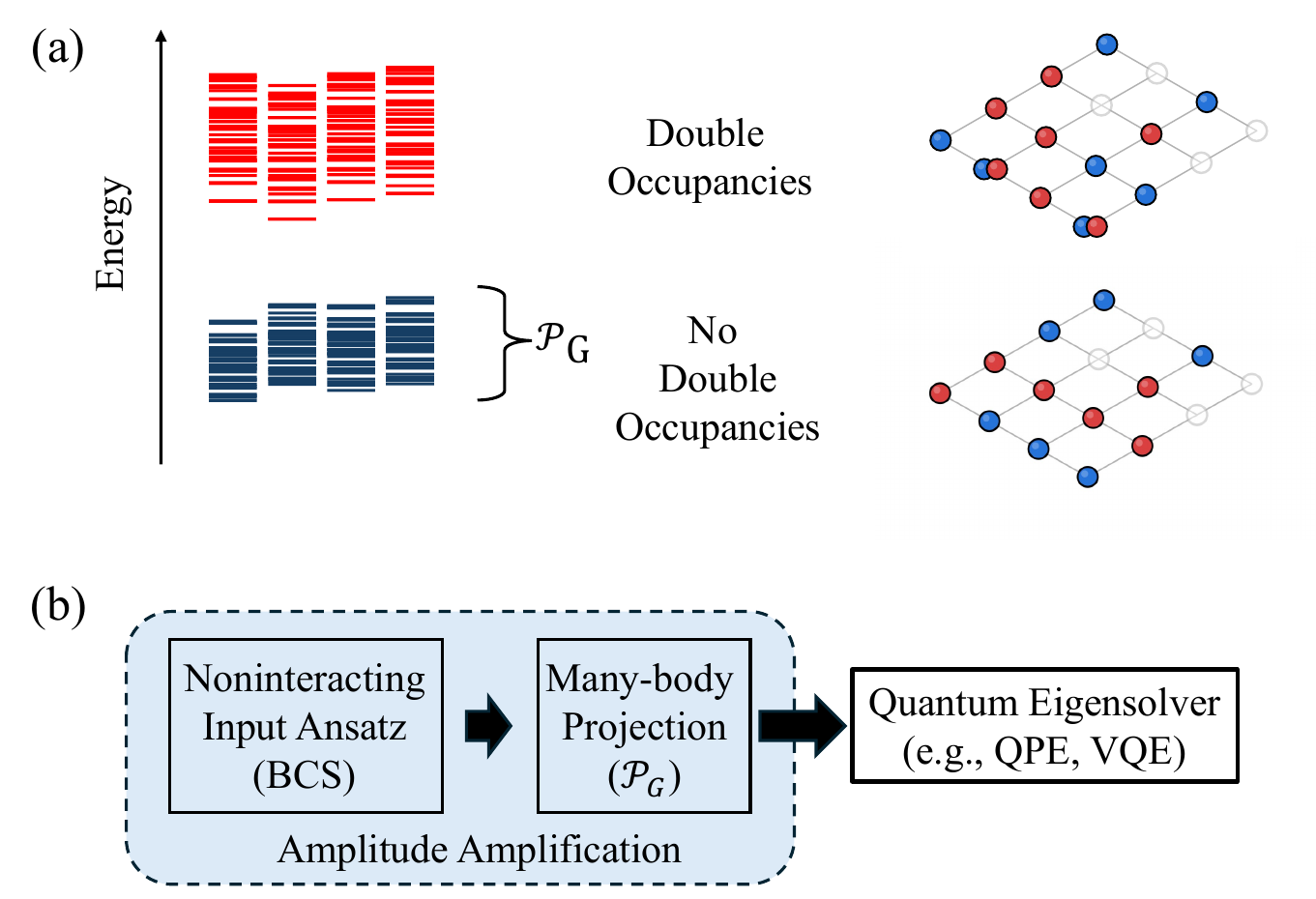}
\caption{{\bf Many-body Hilbert space and workflow schematics}
(a) A schematic of the energy spectra expected from a strongly repulsive lattice model where the lower energy manifold has no double occupancies (blue dashes) and the high energy states have at least one double occupancy (red dashes).  The lattice schematics contain $N_s=16$ sites.  Four states are possible at each site: empty (a hole), a spin up fermion (red sphere), a spin down fermion (blue sphere), and doubly occupied.  $4^{N_s}$ basis configurations superpose to define many-body eigenstates.  Gutzwiller projection excludes doubly occupied basis configurations. (b) Schematic workflow showing a quantum algorithm with an input BCS state parameterized to capture the relevant lattice model parameters (e.g., hopping matrix elements, chemical potential, lattice structure, and disorder).  Amplitude amplification is then used to speed up Gutzwiller projection.
}
\label{fig:schematic}
\end{figure}

Quantum simulation offers a possible route to studying strongly correlated lattice models on system sizes beyond the reach of exact classical methods
\cite{Feynman1982,Georgescu2014,DALEY2022}. Solving these models will, in turn, aid in predicting quantum materials properties and settling debates over the fundamental nature of quantum states of matter. Examples of debated issues include the properties of quantum spin liquid models of magnetism (as captured by spin lattice models) \cite{ZHOU2017a} and the nature of the pairing mechanism in high-temperature superconductors as captured by the Hubbard lattice models \cite{auerbach1994interacting,AROVAS2022}. Hubbard models consist of spinful fermions hopping on a lattice with local (typically on-site) repulsion (see Fig.~\ref{fig:schematic}a) and are believed to harbor key features of high-temperature superconducting compounds, including pairing that arises solely from repulsion \cite{Anderson87}. The Bardeen-Cooper-Schrieffer (BCS) wavefunction \cite{BARDEEN1957} was designed as a paired state for describing weakly interacting models of low-temperature superconductivity. But the BCS ansatz fails to accurately capture the ground state of the strongly repulsive models of high-temperature superconductors.

Quantum simulation algorithms \cite{Georgescu2014} offer a wealth of outputs that can be used to accurately identify ground states and other properties of otherwise intractable strongly correlated models of materials \cite{Ortiz2001,Wecker2015}. For example, variational quantum eigensolvers (VQEs)  \cite{Peruzzo2014,Wecker2015,mcclean2016theory,cerezo2021variational} and quantum phase estimation (QPE) \cite{Kitaev1995,Abrams1999} can be used to obtain energy eigenvalues. Dynamics can be accessed with product formulas \cite{LLOYD1996}, the linear combination of unitaries method \cite{Childs2012c}, and block-encoding methods \cite{MARTYN2021}. However, all of these methods are sensitive to the choice of the input (initial) state. Choosing initial states close to the exact ground state is crucial for building fast and accurate quantum simulation algorithms.

Useful variational initial states are known for some repulsive Hubbard and related spin models in strongly correlated regimes.
Anderson \cite{Anderson87} constructed an ansatz for ground states by Gutzwiller projecting the BCS state [this state is often called the resonating-valence-bond (RVB) state]. The Gutzwiller projection operator, ${\cal P}_G$, acts on all lattice sites simultaneously to remove basis states with double occupancy across the entire lattice \cite{GUTZWILLER1963,GUTZWILLER1964,GUTZWILLER1965}. The BCS state has tunable parameters to accommodate a broad class of models, while ${\cal P}_G$ accounts for strong repulsion. ${\cal P}_G$ is consistent with the Hilbert space structure expected from the strongly repulsive limit (see Fig.~\ref{fig:schematic}a), where, for strong on-site repulsion, we expect a partitioning between low-energy (singly occupied) states and unwanted high-energy (doubly occupied) states.
Small-system numerics~\cite{Park05,Park05_PRB,Kwon22} show that the RVB state can be an accurate ansatz in strongly correlated regimes, motivating the question of whether it can also be prepared efficiently enough to serve as a useful quantum-algorithm input state.

Two critical issues arise when implementing the Gutzwiller/RVB approach as a quantum-algorithm initial state on large system sizes. First, we need an efficient quantum algorithm to construct the BCS state. On finite-size quantum computers, this task can be nontrivial because the resulting BCS state may not preserve translational symmetry. Second, we need an efficient quantum algorithm to perform the Gutzwiller projection on the BCS state. ${\cal P}_G$ can, in principle, be implemented by a method that repeatedly measures double occupancy at every site and post-selects outcomes with no double occupancy~\cite{Murta21, Seki22}; however, this measurement-based approach is highly inefficient because the probability weight of the RVB state within the BCS state, ${\cal W}$, is, as we will show, extremely small. This issue is especially pressing because the real-space RVB state can be parameterized to capture large classes of strongly correlated models that include disorder and a variety of lattice configurations, thus offering a versatile initial state for quantum algorithms on many different models of materials.

To solve this initial-state problem, we introduce and characterize a quantum algorithm dubbed amplitude amplification for Gutzwiller projection (AAGP), Fig.~\ref{fig:schematic}b. We propose that amplitude amplification~\cite{grover1996fast,Yoder14, Gilyen19, MARTYN2021} can coherently increase the weight of the Gutzwiller-projected component of the BCS state, yielding the same quadratic query improvement as Grover-style search to speed up the preparation of RVB states. We show that our algorithm requires ${\cal O}(1/\sqrt{\cal W})$ queries of the BCS state, whereas the measurement-based projection algorithm requires ${\cal O}(1/{\cal W})$ queries. As a result, we find much more favorable scaling for fault-tolerant implementations by showing that the $T$-gate count for the RVB input state on $N_s$ sites is ${\cal O} \left(\frac{\ln (2/\delta)}{\sqrt{\cal W}} N_s^2 \log_2 (1/\epsilon)  \right)$, where $\delta$ and $\epsilon$ are the tolerance and precision, respectively. As an example, we show numerically that using AAGP to prepare the RVB state on the 100-site $t$-$J$ model (a model that is a limiting case of the Hubbard model) requires $\sim 10^7$ fewer queries than the measurement-based projection algorithm.

The significance of this reduction is not merely asymptotic. In the parameter regime benchmarked here, the quadratic improvement translates into many orders of magnitude in the number of required queries and can move projected-state preparation from effectively unusable to plausibly deployable within a larger fault-tolerant workflow. In this sense, the gain is sufficient to cross a practical finite-size threshold even though the overall scaling remains governed by the exponentially small RVB weight. This finite-size threshold is important because system sizes on the order of 100 sites are already well beyond the exact-diagonalization regime used to calibrate the ansatz and the overlap scaling, while also being large enough to probe physically relevant many-body behavior. We therefore do not view AAGP merely as an abstract quadratic improvement, but as an enabling input-state preparation primitive for physically motivated projected BCS/RVB states.

AAGP can be used to prepare input states for many different lattice models. The test case discussed here is the square-lattice $t$-$J$ model. But the quantum algorithm we implement for the BCS ansatz allows a versatile set of input parameters. Translational symmetry is not assumed. Real-space and disordered models may therefore be implemented. The AAGP algorithm also accommodates different lattice geometries, including colored inter-site hopping parameters. The filling (defined as half the number of particles per lattice site) can also be tuned. At half-filling, related projected fermionic states can also serve as useful variational states for some frustrated quantum antiferromagnets \cite{auerbach1994interacting}, including cases where spin-liquid ground states are actively discussed \cite{Gros88,Paramekanti01}. A notable example is the Dirac spin liquid, which is essentially the Gutzwiller-projected Fermi sea filled to the Dirac nodes~\cite{Wen02,Hermele04}. Dirac spin liquids are regarded as one of the most promising candidates for the ground state of frustrated spin-1/2 systems, such as on the kagome lattice~\cite{Ran07,Hermele08}. The AAGP algorithm therefore provides a versatile class of input states whose preparation is substantially more resource-efficient than measurement-based postselection.

The rest of the paper is organized as follows.
Section~\ref{sec:overall} describes the overall structure of the AAGP algorithm.
Section~\ref{sec:BCS} explains how to construct the BCS state via the Bloch-Messiah decomposition.  Section~\ref{sec:amplitude_amplification} explains how to perform the Gutzwiller projection via amplitude amplification.
Section~\ref{sec:RVB_weight} calculates the probability weight of the RVB state within the BCS state to estimate the quantum advantage of the amplitude amplification algorithm.
We conclude in Sec.~\ref{sec:discussions}.

\begin{figure*}[t]
\includegraphics[width=0.9\textwidth]{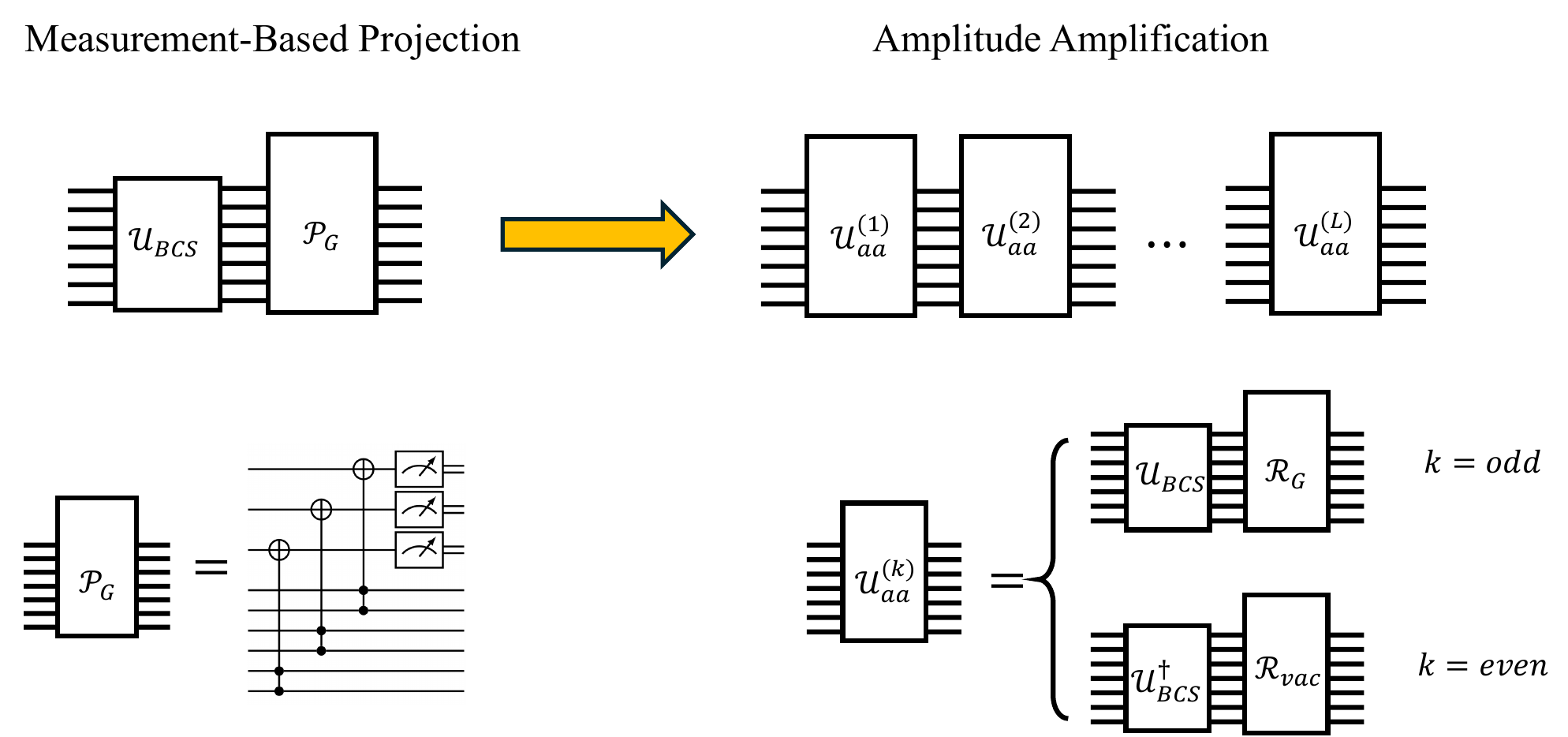}
\caption{{\bf Overall structure of the amplitude amplification algorithm for the Gutzwiller projection.} 
The measurement-based projection algorithm measures double occupancy at every site and postselects the outcome with no double occupancy.
${\cal U}_{\rm BCS}$ is the unitary operator that creates the BCS state from the vacuum or the null state $\vert {\bf 0} \rangle=\vert 0 \rangle^{\otimes N}$. 
${\cal P}_{\rm G}$ implements the Gutzwiller projection by measuring double occupancy at every site using CCNOT gates.
The amplitude amplification algorithm replaces ${\cal P}_{\rm G}$ with a series of unitary operations ${\cal U}^{(k)}_{\rm aa}$ with $k=1, \cdots, L$, where $L$ is determined by two factors: (i) the probability weight of the RVB state within the BCS state, ${\cal W}$, and (ii) the desired tolerance, $\delta$. 
Concretely, $L$ is an odd integer such that $L \ge \ln{(2/\delta)}/\sqrt{{\cal W}}$.
${\cal U}^{(k)}_{\rm aa}$ is given as a product of two unitary operators: (i) ${\cal U}_{\rm BCS} {\cal R}_{\rm G}$ if $k$ is odd, and (ii) ${\cal U}^\dagger_{\rm BCS} {\cal R}_{\rm vac}$ if $k$ is even, where ${\cal R}_{\rm G}$ and ${\cal R}_{\rm vac}$ are the the conditional phase-rotation unitaries associated with the Gutzwiller-projected subspace and the vacuum state, respectively.
} 
\label{fig:Overall_structure}
\end{figure*}

\section{Overall structure}
\label{sec:overall}

The RVB state can be constructed by applying the Gutzwiller projection operator, ${\cal P}_{\rm G}$, to the BCS state:
\begin{equation}
|\psi_{\rm RVB}\rangle = {\cal C} {\cal P}_{\rm G} |\psi_{\rm BCS}\rangle ,
\end{equation}
where ${\cal C}$ is the normalization constant.
In a straightforward approach, $\mathcal{P}_{\mathrm{G}}$ can be implemented by performing projective measurements at every site to remove all configurations with double occupancy.
Specifically, the following quantum circuit can implement $\mathcal{P}_{\mathrm{G}}$ by applying the CCNOT gates to every pair of neighboring qubits representing the spin-up and spin-down states at each site:
\begin{equation}
\begin{quantikz}[row sep={0.3cm,between origins}, column sep=0.3cm]
\lstick[wires=3]{$\ket{0}^{\otimes N_s}$} & \qw & \qw & \targ{} & \meter{} & \cw \rstick[wires=3]{$\boldsymbol{m}$} \\ [0.3cm]
& \qw & \targ{} & \qw & \meter{} & \cw \\ [0.3cm]
& \targ{} & \qw & \qw & \meter{} & \cw \\ [0.3cm]
\lstick[wires=6]{\ket{\Psi_\textrm{BCS}}} & \qw & \qw & \control{} & \qw & \qw \rstick[wires=6]{$| \Psi_{\boldsymbol{m}} \rangle$} \\
& \qw & \qw & \ctrl{-4} & \qw & \qw \\
& \qw & \control{} & \qw & \qw & \qw \\
& \qw & \ctrl{-5} & \qw & \qw & \qw \\
& \control{} & \qw & \qw & \qw & \qw \\
& \ctrl{-6} & \qw & \qw & \qw & \qw
\end{quantikz} ,
\label{eq:Naive_projection}
\end{equation}
where $N_s$ denotes the total number of sites.
When the measurement outcome in the ancilla qubits becomes a null vector, i.e., $\boldsymbol{m} = 0^{\otimes N_s}$, we obtain the RVB state, i.e., $\vert \Psi_{\boldsymbol{m}=0^{N_s}} \rangle \propto \vert \Psi_\textrm{RVB} \rangle$.

The measurement-based projection algorithm is conceptually natural but inefficient in practice.
Because of the destructive nature of the measurement, any measurement "failure" with only a single outcome of 1 leads to the collapse of the entire wave function outside the Gutzwiller-projected Hilbert space.
Thus, whenever postselection fails, the protocol must restart from the initial state, typically the vacuum or the null state $\vert {\bf 0} \rangle=\vert 0 \rangle^{\otimes N}$ ($N=2N_s$), and regenerate the BCS state by applying the unitary operator $\mathcal{U}_\textrm{BCS}$. 
As a result, the measurement-based projection algorithm requires executing $\mathcal{U}_\textrm{BCS}$ a number of times inversely proportional to the probability weight of the RVB state within the BCS state, which is exponentially suppressed as the system size increases. 
In other words, the success probability of the measurement-based projection algorithm is extremely low in any reasonably large systems, as noted in the previous works~\cite{Murta21, Seki22}.

In this context, it is vital to develop an alternative quantum algorithm that can systematically raise the success probability of the Gutzwiller projection close to one at will.
In this work, we employ the recently developed amplitude amplification method~\cite{Yoder14, Gilyen19, MARTYN2021} to construct scalable quantum algorithms for the Gutzwiller projection, which can guarantee an arbitrarily high success probability of the Gutzwiller projection while using significantly fewer $\mathcal{U}_\textrm{BCS}$ and $\mathcal{U}^\dagger_\textrm{BCS}$ than the measurement-based projection algorithm.

Figure~\ref{fig:Overall_structure} shows the overall structure of the AAGP algorithm.
The essence of the amplitude amplification algorithm (or quantum signal processing in general) can be understood in terms of Grover's quantum search algorithm, which provides a quadratic speedup over the measurement-based search. 
Then, the AAGP algorithm can generate the RVB state to within $\delta$ using only ${\cal O}(1/\sqrt{{\cal W}})$ applications of $\mathcal{U}_\textrm{BCS}$ and $\mathcal{U}^\dagger_\textrm{BCS}$, together with the conditional unitaries introduced below.
This provides a quadratic improvement over the measurement-based projection algorithm, which requires ${\cal O}(1/{\cal W})$ applications.
Details of the AAGP algorithm are provided in the sections below.

\section{Construction of the BCS state}
\label{sec:BCS}

In this section, we describe how to prepare an arbitrary BCS state using a unitary quantum circuit. Although related constructions have appeared in the literature~\cite{jiang2018fermions}, we provide a step-by-step derivation because this preparation procedure is a key primitive for the quantum algorithms developed in this work.

The BCS state is the ground state of the BCS Hamiltonian, which can be expressed in real space as the Bogoliubov-de Gennes (BdG) Hamiltonian:
\begin{align}
H_{\rm BdG} &= \sum_{i, j=1}^{N} M_{ij} c_i^\dagger c_j + \frac{1}{2} \sum_{i, j=1}^{N} \big( \Delta_{ij} c_i^\dagger c_j^\dagger + \textrm{H.c.} \big) 
\nonumber \\
&= \frac{1}{2} 
\left( 
\begin{array}{cc}
\mathbf{c}^\dagger & \mathbf{c} 
\end{array} 
\right) 
{\bf H}
\left( 
\begin{array}{c}
\mathbf{c} \\
\mathbf{c}^\dagger
\end{array} 
\right) 
+(\textrm{c-number}) ,
\label{eq:BdG-Ham}
\end{align}
where $\mathbf{c}=(c_1,\cdots,c_N)$ with $N$ being the number of states distinguished by both site and spin indices, i.e., $N=2 N_s$. 
Meanwhile, the single-particle Hamiltonian matrix ${\bf H}$ is given as follows:
\begin{align}
{\bf H}=\left( 
\begin{array}{cc}
\mathbf{M} & \mathbf{\Delta} \\
-\mathbf{\Delta}^* & -\mathbf{M}^*
\end{array} \right) ,
\label{eq:H_definition}
\end{align}
where $\mathbf{M}=[M_{ij}]$ and $\mathbf{\Delta}=[\Delta_{ij}]$ are complex matrices satisfying $\mathbf{M}^\dagger = \mathbf{M}$ and $\mathbf{\Delta}^t = -\mathbf{\Delta}$, which implies that $H_{\rm BdG}$ is Hermitian.
It is important to note that the BdG Hamiltonian in Eq.~\eqref{eq:BdG-Ham} can describe the Dirac Fermi sea state in the kagome lattice by choosing $\mathbf{M}$ appropriately while setting $\mathbf{\Delta}$ to zero.

We can obtain the BCS state by exactly diagonalizing ${\bf H}$:
\begin{equation}
\mathbf{W}^\dagger {\bf H} \mathbf{W}
= \textrm{diag} \left( \epsilon_1, \cdots, \epsilon_N, -\epsilon_1, \cdots, -\epsilon_N \right),
\label{eq:W_definition}
\end{equation}
with
\begin{equation}
\mathbf{W} = 
\left( \begin{array}{cc}
{\bf U} & {\bf V}^* \\
{\bf V} & {\bf U}^*
\end{array} \right)
\label{eq:UV_definition} 
\end{equation}
where ${\bf U}$ and ${\bf V}$ are two $N \times N$ matrices such that
\begin{align}
&{\bf U}^\dagger {\bf U} + {\bf V}^\dagger {\bf V} = {\bf 1}_{N \times N}, \;\;\;\; {\bf U} {\bf U}^\dagger + {\bf V}^* {\bf V}^t = {\bf 1}_{N \times N}, \nonumber \\
&{\bf U}^t {\bf V} + {\bf V}^t {\bf U} = {\bf 0}_{N \times N}, \;\;\;\; {\bf U} {\bf V}^\dagger + {\bf V}^* {\bf U}^t = {\bf 0}_{N \times N},
\label{eq:UV_identities}
\end{align}
which implies that ${\bf W}$ is unitary.
Note that the eigenvalues always appear in pairs with opposite signs. 

Performing exact diagonalization of the single-particle Hamiltonian matrix ${\bf H}$ in Eq.~\eqref{eq:W_definition} is not too difficult for finite-size systems because the matrix size of ${\bf H}$ is only four times the total number of sites.
Therefore, we can perform exact diagonalization of ${\bf H}$ on a classical computer and then feed the resulting information into the quantum algorithm to construct the BCS state on a quantum computer.

After exact diagonalization of ${\bf H}$, the BdG Hamiltonian can be written as
\begin{equation}
H_{\rm BdG} = \sum_i^N \epsilon_i \gamma^\dagger_i \gamma_i +(\textrm{c-number}),
\label{eq:BdG_Ham_diagonalized}
\end{equation}
which gives rise to the BCS state as the unique ground state annihilated by all Bogoliubov quasiparticle annihilation operators $\gamma_i$ with $i \in (1,\cdots,N)$.
Bogoliubov quasiparticle operators, $\boldsymbol{\gamma}=(\gamma_1,\cdots,\gamma_N)$ and $\boldsymbol{\gamma}^\dagger=(\gamma_1^\dagger,\cdots,\gamma_N^\dagger)$, are related to electronic counterparts via
\begin{align}
\boldsymbol{\gamma}&=\mathbf{U}^\dagger \mathbf{c}+\mathbf{V}^\dagger \mathbf{c}^\dagger,
\nonumber \\
\boldsymbol{\gamma}^\dagger&=\mathbf{V}^t \mathbf{c}+\mathbf{U}^t \mathbf{c}^\dagger.
\end{align}
Now, a crucial question is how to construct the BCS state in terms of the $\mathbf{c}$ operators.

\subsection{Bloch-Messiah Decomposition}

The BCS state can be constructed by using the Bloch-Messiah decomposition~\cite{Bloch62,Ring80}.
Specifically, the BCS state can be expressed as follows:
\begin{equation}
|\Psi_{\rm BCS}\rangle = \prod_{i=1}^n f^\dagger_i \prod_{j=1}^p \left( u_j +v_j f^\dagger_{n+2j-1} f^\dagger_{n+2j} \right) |{\rm vac}\rangle,
\label{eq:BCS}
\end{equation}
where $|{\rm vac}\rangle$ indicates the vacuum state void of any electrons.
New fermion operators, $\mathbf{f}=(f_1,\cdots,f_N)$ and $\mathbf{f}^\dagger=(f_1^\dagger,\cdots,f_N^\dagger)$, are related to electronic counterparts via
\begin{align}
\mathbf{f}=\mathbf{D}^\dagger \mathbf{c}, \;\;\;\;
\mathbf{f}^\dagger=\mathbf{D}^t \mathbf{c}^\dagger,
\end{align}
where $\mathbf{D}$ is the unitary matrix diagonalizing the Hermitian matrix $\boldsymbol{\rho}=\mathbf{1}_{N \times N}-\mathbf{U}\mathbf{U}^\dagger$ such that
\begin{equation}
\mathbf{D}^\dagger \boldsymbol{\rho} \mathbf{D} 
= \left( \begin{array}{ccc}
\cline{1-1}
\lrvertline{\mathbf{1}_{n \times n}} & & \\
\cline{1-2}
& \lrvertline{\bigoplus\limits_{j=1}^{p} v_j^2 \boldsymbol{\sigma}_0} & \\
\cline{2-3}
& & \lrvertline{\mathbf{0}_{m \times m}} \\
\cline{3-3}
\end{array} \right),
\end{equation}
where $\boldsymbol{\sigma}_0$ indicates the zeroth Pauli matrix. 
Note that the eigenvalues of $\boldsymbol{\rho}$ consist of $n$ ones, $p$ pairs of $v^2_j$, and $m$ zeros, so that $N=n+2p+m$.
Physically, $n$ represents the number of unpaired occupied states, $2p$ the number of paired states, and $m$ the number of unoccupied states.
When momentum is well-defined, $\mathbf{D}$ simply corresponds to the Fourier transform from ${\bf c}$ to ${\bf f}$, in which case Eq.~\eqref{eq:BCS} reduces to the standard form of the BCS state written in momentum space.
However, keep in mind that we aim to construct the BCS state on finite-size quantum computers that may not necessarily preserve translational symmetry. 
To this end, we need a general protocol to transform from $\mathbf{c}$ to $\mathbf{f}$.

To begin, if $\mathbf{D}$ were the identity matrix, the BCS state can be generated using a depth-1 quantum circuit:
\begin{equation}
|\Psi_{\rm BCS}^\textrm{depth-1}\rangle = \prod_{i=1}^n c^\dagger_i \prod_{j=1}^p \left( u_j +v_j c^\dagger_{n+2j-1} c^\dagger_{n+2j} \right) |{\rm vac}\rangle,
\label{eq:BCS_depth-1}
\end{equation}
which requires only single-qubit operations for unpaired occupied states and two-qubit operations for paired states.
Of course, $\mathbf{D}$ is generally not the identity matrix, and therefore the depth-1 BCS state is not the full BCS state that diagonalizes the BdG Hamiltonian.
Fortunately, the full BCS state in Eq.~\eqref{eq:BCS} can be obtained from the depth-1 BCS state by applying a series of appropriate unitary operators that transform from $\mathbf{c}$ to $\mathbf{f}$.

Transforming from $\mathbf{c}$ to $\mathbf{f}$ essentially amounts to diagonalizing $\mathbf{D}$ via Givens rotations.
To be more precise, we need to diagonalize the first $n+2p$ columns of the $N \times N$ matrix $\mathbf{D}$. 
Note that only the fermionic modes with indices from $1$ to $n+2p$ are used to construct the BCS state.
In summary, denoting the first $n+2p$ columns of $\mathbf{D}$ as $\mathbf{D}|_{n+2p}$, our task is to perform the singular value decomposition of $\mathbf{D}|_{n+2p}$ by using an appropriate series of Givens rotations.

Givens rotations can be expressed using the following matrix:
\begin{equation}
\mathbf{G}_{j,j+1} (\theta, \phi) \equiv
\left( 
\begin{array}{ccccc}
1 &  &  &  &  \\ 
& \ddots & & & \\ \cline{3-3}
& & \lrvertline{\mathbf{G}(\theta, \phi)} & & \\ \cline{3-3}
& & & \ddots &  \\
& & & & 1
\end{array}
\right),
\label{eq:Givens}
\end{equation}
where
\begin{equation}
\mathbf{G}(\theta, \phi) \equiv
\left( \begin{array}{cc}
e^{i \phi} \cos \theta & -e^{i \phi} \sin \theta \\
e^{-i \phi} \sin \theta & e^{-i \phi} \cos \theta
\end{array} \right).
\end{equation}
is located in the $2 \times 2$ diagonal block containing the $j$ and $(j+1)$-th diagonal elements.

Before applying the Givens rotations, however, it is useful to simplify $\mathbf{D}|_{n+2p}$, which can reduce the number of Givens rotations needed for the singular value decomposition.
This simplification is possible because there is a redundancy in the form of $\mathbf{D}|_{n+2p}$ that can result in the same BCS state.
We can use this redundancy to make $[\mathbf{D}|_{n+2p}]_{ij}=0$ for $i - j > N - n$. 
Please refer to the Appendix~\ref{appen:Redundancy} for more information.

It is helpful to understand this redundancy and the subsequent Givens rotations through a simple example.
From this forward, let us consider an example with $N=6$, $n=3$, and $p=1$, in which case $\mathbf{D}|_{n+2p=5}$ can be simplified as
\begin{align}
\mathbf{D}|_{5}=
\left( 
\begin{array}{ccc|cc}
* & * & * & * & * \\
* & * & * & * & * \\
* & * & * & * & * \\
* & * & * & * & * \\
0 & * & * & * & * \\
0 & 0 & * & * & * 
\end{array}
\right),
\end{align}
where $*$ indicates some arbitrary numbers, and the vertical line separates the $n$-th and $(n+1)$-th columns.
As can be seen, the bottom-left corner of the matrix becomes zero.
Note that this simplification is performed at the preparation stage before computing all the necessary Givens rotation angles to construct the BCS state in a quantum circuit.

Next, we need to compute the necessary Givens rotation angles to perform the singular value decomposition of $\mathbf{D}|_{n+2p}$.
Let us apply  $\mathbf{G}_{3,4} (\theta_1, \phi_1)$ to $\mathbf{D}|_{5}$ as follows:
\begin{align}
\mathbf{G}_{3,4} (\theta_1, \phi_1) \mathbf{D}|_{5}
= \left( \begin{array}{ccc|cc}
* & * & * & * & * \\ 
* & * & * & * & * \\ \cline{1-2}
\lvertline{*} & \rvertline{*} & * & * & * \\ 
\lvertline{{\bf 0}} & \rvertline{*} & * & * & * \\ \cline{1-2}
0 & * & * & * & * \\ 
0 & 0 & * & * & * 
\end{array} 
\right),
\label{eq:Givens_rotation1}
\end{align}
where $\theta_1$ and $\phi_1$ are chosen so that the bottom-left element (bold-faced zero) in the $2 \times 2$ box of the right-hand matrix becomes zero.
The specific values of $\theta_1$ and $\phi_1$ can be computed in advance on a classical computer and then fed into the quantum algorithm.

The main idea is to repeatedly perform similar Given rotations until the resulting matrix is fully singular-value-decomposed.
Again, let us take the previous example and perform the following series of Givens rotations.
\begin{widetext}
\begin{align}
\left( \begin{array}{ccc|cc}
* & * & * & * & * \\ 
* & * & * & * & * \\ \cline{1-2}
\lvertline{*} & \rvertline{*} & * & * & * \\ 
\lvertline{{\bf 0}} & \rvertline{*} & * & * & * \\ \cline{1-2}
0 & * & * & * & * \\ 
0 & 0 & * & * & * 
\end{array} \right) 
&\xmapsto{\mathbf{G}_{2,3}(2) \mathbf{G}_{4, 5}(3)}  
\left( \begin{array}{ccc|cc}
* & * & * & * & * \\ \cline{1-2}
\lvertline{*} & \rvertline{*} & * & * & * \\ 
\lvertline{{\bf 0}} & \rvertline{*} & * & * & * \\ \cline{1-3}
0 & \lvertline{*} & \rvertline{*} & * & * \\ 
0 & \lvertline{{\bf 0}} & \rvertline{*} & * & * \\ \cline{2-3}
0 & 0 & * & * & * 
\end{array} \right)
\xmapsto{\mathbf{G}_{1,2} (4) \mathbf{G}_{3,4}(5) \mathbf{G}_{5,6}(6)} 
\left( \begin{array}{ccc|cc} \cline{1-2}
\lvertline{\lambda_1} & \rvertline{\underline{0}} & \underline{0} & \underline{0} & \underline{0} \\
\lvertline{{\bf 0}} & \rvertline{*} & * & * & * \\ \cline{1-3}
0 & \lvertline{*} & \rvertline{*} & * & * \\ 
0 & \lvertline{{\bf 0}} & \rvertline{*} & * & * \\ \cline{2-4}
0 & 0 & \lvertline{*} & \lrvertline{*} & * \\ 
0 & 0 & \lvertline{{\bf 0}} & \lrvertline{*} & * \\ \cline{3-4}
\end{array} \right) 
\nonumber \\
&\xmapsto{\mathbf{G}_{2,3}(7) \mathbf{G}_{4,5}(8)} 
\left( \begin{array}{ccc|cc}
\lambda_1 & 0 & 0 & 0 & 0 \\ \cline{2-3}
0 & \lvertline{\lambda_2} & \rvertline{\underline{0}} & \underline{0} & \underline{0} \\ 
0 & \lvertline{{\bf 0}} & \rvertline{*} & * & * \\ \cline{2-4}
0 & 0 & \lvertline{*}  & \lrvertline{*} & * \\ 
0 & 0 & \lvertline{{\bf 0}} & \lrvertline{*} & * \\ \cline{3-4}
0 & 0 & 0 & * & * 
\end{array} \right) 
\xmapsto{\mathbf{G}_{3,4}(9) \mathbf{G}_{5,6}(10)} 
\left( \begin{array}{ccc|cc}
\lambda_1 & 0 & 0 & 0 & 0 \\ 
0 & \lambda_2 & 0 & 0 & 0 \\ \cline{3-4}
0 & 0 & \lvertline{\lambda_3} & \lrvertline{\underline{0}} & \underline{0} \\ 
0 & 0 & \lvertline{{\bf 0}}  & \lrvertline{*} & * \\ \cline{3-5}
0 & 0 & 0 & \lvertline{*} & \rvertline{*} \\ 
0 & 0 & 0 & \lvertline{{\bf 0}} & \rvertline{*} \\ \cline{4-5}
\end{array} \right) \nonumber \\
&\xmapsto{\mathbf{G}_{4,5}(11)} 
\left( \begin{array}{ccc|cc}
\lambda_1 & 0 & 0 & 0 & 0 \\ 
0 & \lambda_2 & 0 & 0 & 0 \\ 
0 & 0 & \lambda_3 & 0 & 0 \\ \cline{4-5}
0 & 0 & 0  & \lvertline{\lambda_4} & \rvertline{\underline{0}} \\ 
0 & 0 & 0 & \lvertline{{\bf 0}} & \rvertline{*} \\ \cline{4-5}
0 & 0 & 0 & 0 & * 
\end{array} \right) 
\xmapsto{\mathbf{G}_{5,6}(12)} 
\left( \begin{array}{ccc|cc}
\lambda_1 & 0 & 0 & 0 & 0 \\ 
0 & \lambda_2 & 0 & 0 & 0 \\ 
0 & 0 & \lambda_3 & 0 & 0 \\ 
0 & 0 & 0  & \lambda_4 & 0 \\ \cline{4-5}
0 & 0 & 0 & \lvertline{0} & \rvertline{\lambda_5} \\
0 & 0 & 0 & \lvertline{0} & \rvertline{{\bf 0}} \\ \cline{4-5}
\end{array} \right) ,
\label{eq:Givens_rotation2}
\end{align}
\end{widetext}
where $\mathbf{G}_{j, j+1}(i)$ is the abbreviation of $\mathbf{G}_{j, j+1} (\theta_i, \phi_i)$ with $\theta_i$ and $\phi_i$ being the Givens rotation angles, ${\bf 0}$ is the zero element vanishing due to the suitable choice of Givens rotation angles, $\underline{0}$ is the zero element vanishing due to the orthogonality between columns, and $\lambda_i$ is a $\mathrm{U}(1)$ complex phase. 
Again, the specific values of $\theta_i$ and $\phi_i$ can be computed in advance on a classical computer and then fed into the quantum algorithm.

The above series of Givens rotations provides the necessary sequence of unitary transformations mapping from the full BCS state in Eq.~\eqref{eq:BCS} to the depth-1 BCS state written as
\begin{align}
&|\Psi_{\rm BCS}^\textrm{depth-1}\rangle 
\nonumber \\
&= \prod_{i=1}^n \lambda_i c^\dagger_i
\prod_{j=1}^p \left( u_j +v_j \lambda_{n+2j-1}\lambda_{n+2j} c^\dagger_{n+2j-1} c^\dagger_{n+2j} \right) |{\rm vac}\rangle,
\label{eq:BCS_depth-1_v2}
\end{align}
which is slightly different from the simpler depth-1 BCS state in Eq.~\eqref{eq:BCS_depth-1} to account for the effects of the phase factors $\lambda_i$'s. 

Finally, our objective is to obtain the full BCS state, starting from the depth-1 BCS state. 
This can be accomplished by reversing the series of Givens rotations described so far.

\subsection{Quantum circuits for the Bloch-Messiah decomposition}

Using the Jordan-Wigner (JW) transformation, we can construct quantum circuits for the depth-1 BCS state and the necessary series of Givens rotations to obtain the full BCS state.

First, apart from the unimportant overall phase factor, the depth-1 BCS state in Eq.~\eqref{eq:BCS_depth-1_v2} can be expressed using the JW transformation as follows:
\begin{equation}
|\Psi_{\rm BCS}^\textrm{depth-1}\rangle \xmapsto{\rm JW}  \prod_{i=1}^n \hat{X}_i \prod_{j=1}^p \hat{F}_{n+2j-1,n+2j}(\theta_j,\phi_j) |0^N\rangle,
\label{eq:BCS_depth-1_v2_JW}
\end{equation}
where $\hat{X}_i$ is the Pauli X operator acting on the $i$-th qubit to create an occupied state, and $\hat{F}_{n+2j-1,n+2j}(\theta_j,\phi_j)$ is the local two-qubit operator acting on the neighboring $n+2j-1$ and $n+2j$ qubits to create a paired state, which can be implemented using the following quantum circuit:
\begin{widetext}
\begin{equation}
\begin{quantikz}
n+2j-1 \; & \gate[2]{\hat{F}(\theta_j, \phi_j)} & \; n+2j-1 \qw \\
n+2j \; & & \; n+2j \qw
\end{quantikz} 
= 
\begin{quantikz}
\qw n+2j-1 \; & \qw & \targ{} & \ctrl{1} & \ctrl{1} & \targ{} & \qw & \; n+2j-1 \qw \\
\qw n+2j \; & \gate{X} & \ctrl{-1} & \gate{e^{i \theta_j \boldsymbol{\sigma}_y}} & \gate{e^{i \phi_j \boldsymbol{\sigma}_z}} & \ctrl{-1} & \gate{X} & \; n+2j \qw
\end{quantikz}  ,
\end{equation}
\label{eq:F_quantum_circuit}
\end{widetext}
where $\theta_j = \tan^{-1}(|v_j^\prime|/|u_j|)$ and $\phi_j = \textrm{arg} (v_j^\prime/u_j)/2$ with $v_j^\prime=v_j\lambda_{n+2j-1}\lambda_{n+2j}$.

Second, the Givens rotation between qubits $k$ and $k+1$, denoted as $\hat{G}_{k,k+1}(\theta_l,\phi_l)$, can be implemented using the following quantum circuit:
\begin{widetext}
\begin{equation}
\begin{quantikz}
k \; & \gate[2]{\hat{G}(\theta_l, \phi_l)} & k \; \qw \\
k+1 \; & & \; k+1 \qw
\end{quantikz}
=
\begin{quantikz}
\qw k \;  & \targ{} & \ctrl{1} & \ctrl{1} & \targ{} & \; k \qw \\
\qw k+1 \; & \ctrl{-1} & \gate{e^{-i \theta_l \boldsymbol{\sigma}_y}} & \gate{e^{i \phi_l \boldsymbol{\sigma}_z}} & \ctrl{-1} & \; k+1 \qw
\end{quantikz},
\label{eq:G_quantum_circuit}
\end{equation}
\end{widetext}
where $\theta_l$ and $\phi_l$ are the Givens rotation angles obtained using the procedure explained in Eqs.~\eqref{eq:Givens_rotation1} and \eqref{eq:Givens_rotation2}.

Finally, combining the quantum circuits for the depth-1 BCS state with the subsequent Givens rotations yields the full BCS state.
To be specific, the quantum circuit for the full BCS state in the preceding example with $N=6$, $n=3$, and $p=1$ can be expressed as follows:
\begin{widetext}
\begin{equation}
\begin{quantikz}[row sep={1cm,between origins}, column sep=0.3cm]
\lstick[wires=3]{$\ket{0}^{\otimes n}$} & \gate{X} \qw  & \qw & \qw & \qw & \qw & \gate[2]{\;\hat{G}(4)\;} & \qw & \qw & \qw \rstick[wires=6]{$\ket{\Psi_{\rm BCS}}$,} \\
& \gate{X} \qw & \qw & \qw & \qw & \gate[2]{\;\hat{G}(7)\;} & & \gate[2]{\;\hat{G}(2)\;} & \qw & \qw \\
& \gate{X} \qw & \qw & \qw & \gate[2]{\;\hat{G}(9)\;} & & \gate[2]{\;\hat{G}(5)\;} & & \gate[2]{\;\hat{G}(1)\;} & \qw \\
\lstick[wires=2]{$\ket{0}^{\otimes 2p}$} & \gate[2]{\;\hat{F}(1)\;} & \qw & \gate[2]{\hat{G}(11)} & & \gate[2]{\;\hat{G}(8)\;} & & \gate[2]{\;\hat{G}(3)\;} & & \qw \\
& & \gate[2]{\hat{G}(12)} & & \gate[2]{\hat{G}(10)} & & \gate[2]{\;\hat{G}(6)\;} & \qw & \qw & \qw \\
\lstick[wires=2]{$\ket{0}^{\otimes m}$} & \qw & & \qw & & \qw & \qw & \qw & \qw & \qw
\end{quantikz}
\end{equation}
\end{widetext}
where $\hat{F}(j)$ and $\hat{G}(l)$ are abbreviations for $\hat{F}(\theta_j,\phi_j)$ and $\hat{G}(\theta_l,\phi_l)$, respectively. 
For convenience, we will now refer to the entire quantum circuit described above as $\mathcal{U}_{\rm BCS}$.

It is straightforward to generalize the above procedure to compute the total number of two-qubit gates and the circuit depth for $\mathcal{U}_{\rm BCS}$ in general situations, both of which are important for estimating the quantum resources required for our AAGP algorithm.
Relegating the detailed derivation to the Appendix, the total number of two-qubit gates is $(N-n)(n+2p) - 2p^2$, and the circuit depth is $N+2p$.
See Appendix~\ref{appen:Two-qubit_counting} for details.

\section{Amplitude amplification for the Gutzwiller projection}
\label{sec:amplitude_amplification}

The amplitude amplification algorithm~\cite{Yoder14, Gilyen19, MARTYN2021} is basically an extension of Grover's quantum search algorithm~\cite{grover1996fast}.
Here, we review key aspects of the amplitude amplification algorithm.
Suppose that we have a unitary operator $\mathcal{U}$ that creates a state with a finite overlap with the target state $\vert \Phi \rangle$ from the vacuum state $\vert {\bf 0} \rangle=\vert 0 \rangle^{\otimes N}$:
\begin{equation}
\mathcal{U} \vert {\bf 0} \rangle = \sqrt{{\cal W}} \vert \Phi \rangle + \sqrt{1-{\cal W}} \vert \Phi^\perp \rangle,
\end{equation}
where ${\cal W}$ represents the probability weight of $|\Phi\rangle$ within the $\mathcal{U}$-operated vacuum state and $|\Phi^\perp\rangle$ is the remaining component perpendicular to $|\Phi\rangle$.
In our AAGP algorithm, $\mathcal{U} = \mathcal{U}_{\rm BCS}$ and $\vert \Phi \rangle = \vert \psi_{\rm RVB} \rangle$:
\begin{equation}
\mathcal{U}_{\rm BCS} \vert {\bf 0} \rangle = \sqrt{{\cal W}} \vert \psi_{\rm RVB} \rangle + \sqrt{1-{\cal W}} \vert \psi^\perp_{\rm RVB} \rangle.
\label{eq:U_BCS_on_vac}
\end{equation}
where ${\cal W}$ now specifies the probability weight of the RVB state within the BCS state and  $|\psi^\perp_{\rm RVB}\rangle$ is the remaining component perpendicular to $\psi_{\rm RVB}$ outside the Gutzwiller-projected Hilbert space, i.e., $|\psi^\perp_{\rm RVB}\rangle \propto (1-{\cal P}_{\rm G})|\psi_{\rm BCS}\rangle$.

Our AAGP algorithm systematically amplifies the amplitude of the RVB state using a series of conditional unitary operators for phase rotation that act on the vacuum state and the RVB state, along with $\mathcal{U}_{\rm BCS}$ and $\mathcal{U}^\dagger_{\rm BCS}$. 
Specifically, the conditional unitary operators acting on the vacuum state and the RVB state are defined as follows:
\begin{align}
\mathcal{R}_{\rm vac} (\varphi) &= \openone + \big( e^{i \varphi} -1 \big) \vert {\bf 0} \rangle \langle {\bf 0} \vert ,
\nonumber \\
\mathcal{R}_{\rm RVB} (\varphi) &= \openone + \big( e^{i \varphi} -1 \big) \vert \psi_{\rm RVB} \rangle \langle \psi_{\rm RVB} \vert ,
\label{eq:conditional_rotation}
\end{align}
where the rotation angle $\varphi$ is to be determined by the formula derived later in this paper.
When $\varphi$ equals $\pi$, the conditional unitary operators reduce to Grover's reflection operators.
The amplitude amplification algorithm adjusts the value of $\varphi$ to reach the target state as efficiently as possible.

Now, the four $N$-qubit unitary operators $\mathcal{U}_{\rm BCS}$, $\mathcal{U}^\dagger_{\rm BCS}$, $\mathcal{R}_{\rm vac} (\varphi)$, and $\mathcal{R}_{\rm RVB} (\varphi)$ can be mapped to their corresponding effective single-qubit operators.
This mapping is known as qubitization.
To understand how qubitization works, let us introduce an orthonormal state to the vacuum state, $\vert {\bf 0}^\perp \rangle$, defined by the following condition:
\begin{equation}
\mathcal{U}_{\rm BCS} \vert {\bf 0}^{\perp} \rangle = \sqrt{1-{\cal W}} \vert \psi_{\rm RVB} \rangle - \sqrt{{\cal W}} \vert \psi^\perp_{\rm RVB} \rangle.
\label{eq:U_BCS_on_vac_perp}
\end{equation}
Combining equations \eqref{eq:U_BCS_on_vac} and \eqref{eq:U_BCS_on_vac_perp} gives rise to:
\begin{equation}
\mathcal{U}_{\rm BCS}
\left( \begin{array}{c}
\vert {\bf 0} \rangle \\
\vert {\bf 0}^\perp \rangle 
\end{array} \right) = \left(
\begin{array}{cc}
\sqrt{{\cal W}} & \sqrt{1-{\cal W}} \\
\sqrt{1-{\cal W}} & -\sqrt{{\cal W}} 
\end{array} \right) \left( \begin{array}{c}
\vert \psi_{\rm RVB} \rangle \\
\vert \psi^\perp_{\rm RVB} \rangle
\end{array} \right),
\end{equation}
and its inverse relation,
\begin{equation}
\mathcal{U}^\dagger_{\rm BCS}
\left( \begin{array}{c}
\vert \psi_{\rm RVB} \rangle \\
\vert \psi^\perp_{\rm RVB} \rangle 
\end{array} \right) = 
\left(\begin{array}{cc}
\sqrt{{\cal W}} & \sqrt{1-{\cal W}} \\
\sqrt{1-{\cal W}} & -\sqrt{{\cal W}} 
\end{array} \right) \left( \begin{array}{c}
\vert {\bf 0} \rangle \\
\vert {\bf 0}^\perp \rangle
\end{array} \right).
\end{equation}
Note that both $\mathcal{U}_{\rm BCS}$ and $\mathcal{U}^\dagger_{\rm BCS}$ can be represented by the same $2 \times 2$ matrix $\mathbf{U}({\cal W})$:
\begin{equation}
\mathbf{U}({\cal W}) \equiv 
\left( \begin{array}{cc}
\sqrt{{\cal W}} & \sqrt{1-{\cal W}} \\
\sqrt{1-{\cal W}} & -\sqrt{{\cal W}} 
\end{array} \right).
\end{equation}

Meanwhile, $\mathcal{R}_{\rm vac}(\varphi)$ operate on $\vert {\bf 0} \rangle$ and $\vert {\bf 0}^\perp \rangle$ as follows:
\begin{equation}
\mathcal{R}_{\rm vac}(\varphi)
\left( \begin{array}{c}
\vert {\bf 0} \rangle \\
\vert {\bf 0}^\perp \rangle 
\end{array} \right) = 
\left( \begin{array}{cc}
e^{i\varphi} & 0 \\
0 & 1 
\end{array} \right) 
\left( \begin{array}{c}
\vert {\bf 0} \rangle \\
\vert {\bf 0}^\perp \rangle
\end{array} \right).
\end{equation}
Similarly, $\mathcal{R}_{\rm RVB}(\varphi)$ operate on $\vert \psi_{\rm RVB} \rangle$ and $\vert \psi_{\rm RVB}^\perp \rangle$ as follows:
\begin{equation}
\mathcal{R}_{\rm RVB}(\varphi)
\left( \begin{array}{c}
\vert \psi_{\rm RVB} \rangle \\
\vert \psi^\perp_{\rm RVB} \rangle 
\end{array} \right) = \left(
\begin{array}{cc}
e^{i\varphi} & 0 \\
0 & 1 
\end{array} \right) 
\left( \begin{array}{c}
\vert \psi_{\rm RVB} \rangle \\
\vert \psi^\perp_{\rm RVB} \rangle
\end{array} \right).
\end{equation}
Again, note that both $\mathcal{R}_{\rm vac}(\varphi)$ and $\mathcal{R}_{\rm RVB}(\varphi)$ can be represented by the same $2 \times 2$ matrix $\mathbf{R}(\varphi)$:
\begin{equation}
\mathbf{R}(\varphi) \equiv 
\left( \begin{array}{cc}
e^{i \varphi} & 0 \\
0 & 1
\end{array} \right).
\end{equation}

Next, let us consider applying $\mathcal{U}_{\rm BCS}$ to $\vert {\bf 0} \rangle$ and $\vert {\bf 0}^\perp \rangle$, and then $\mathcal{R}_{\rm RVB}(\varphi_1)$ to the resulting states, which corresponds to the following multiplication:
\begin{align}
\left( \begin{array}{c}
\vert \psi_1 \rangle \\
\vert \psi_1^\perp \rangle 
\end{array} \right) 
&=\mathcal{R}_{\rm RVB}(\varphi_1)
\mathcal{U}_{\rm BCS}
\left( \begin{array}{c}
\vert {\bf 0} \rangle \\
\vert {\bf 0}^\perp \rangle 
\end{array} \right) 
\nonumber \\
&= 
\mathbf{R}(\varphi_1)
\mathbf{U}({\cal W})
\left( \begin{array}{c}
\vert \psi_{\rm RVB} \rangle \\
\vert \psi^\perp_{\rm RVB} \rangle
\end{array} \right),
\end{align}
where $\varphi_1$ is to be determined by the formula derived later in the paper. 
This is the first step in our series of unitary operations for the amplitude amplification algorithm.

The second step involves applying $\mathcal{U}_{\rm BCS}^\dagger$ and then $\mathcal{R}_{\rm vac}(\varphi_2)$ to the states obtained from the first step:
\begin{align}
\left( \begin{array}{c}
\vert \psi_2 \rangle \\
\vert \psi_2^\perp \rangle 
\end{array} \right) 
&=\mathcal{R}_{\rm vac}(\varphi_2)
\mathcal{U}_{\rm BCS}^\dagger
\left( \begin{array}{c}
\vert \psi_1 \rangle \\
\vert \psi_1^\perp \rangle 
\end{array} \right) 
\nonumber \\
&= 
\mathbf{R}(\varphi_2)
\mathbf{U}({\cal W})
\mathbf{R}(\varphi_1)
\mathbf{U}({\cal W})
\left( \begin{array}{c}
\vert {\bf 0} \rangle \\
\vert {\bf 0}^\perp \rangle
\end{array} \right),
\end{align}
where $\varphi_2$ is again to be determined by the formula derived later in the paper. 
Afterward, one can repeat these steps multiple times.

Specifically, after the $n$-th step, the resulting states are given by
\begin{align}
\vert \psi_n \rangle = 
\begin{cases} 
\alpha_n \vert {\bf 0} \rangle  + \beta_n \vert {\bf 0}^\perp \rangle \qquad \qquad \;\; \textrm{if $n =$ even}, \\
\alpha_n \vert \psi_{\rm RVB} \rangle  + \beta_n \vert \psi_{\rm RVB}^\perp \rangle \qquad \textrm{if $n =$ odd},
\end{cases}
\end{align}
where $\alpha_n$ and $\beta_n$ satisfy the following recursion relation:
\begin{align}
\left( \begin{array}{c}
\alpha_n \\
\beta_n 
\end{array} \right) 
=\mathbf{R}(\varphi_n)
\mathbf{U}({\cal W})
\left( \begin{array}{c}
\alpha_{n-1} \\
\beta_{n-1}
\end{array} \right),
\end{align}
with the initial condition $(\alpha_0,\beta_0)=(1,0)$.
Note that, for odd $n$, $|\alpha_n|^2$ indicates the success probability of obtaining the RVB state after the $n$-th step of the iteration.
Similarly, for odd $n$, $|\beta_n|^2=1-|\alpha_n|^2$ indicates the failure probability of not obtaining the RVB state after the $n$-th step of the iteration.

After doing some simple calculations, the $\beta$ recursion relation can be derived as follows:
\begin{align}
\beta_n = \sqrt{{\cal W}} \big(e^{i \varphi_{n-1}} -1\big) \beta_{n-1} + e^{i \varphi_{n-1}} \beta_{n-2},
\label{eq:Beta-recursion}
\end{align}
where $(\beta_0, \beta_1) = (0, \sqrt{1-{\cal W}})$. 
The final state obtained from the iteration process is $\alpha_L \vert \psi_{\rm RVB} \rangle + \beta_L \vert \psi^\perp_{\rm RVB} \rangle$ for some odd $L$, in which case the failure probability for the RVB state is given as $|\beta_L|^2$. 
The amplitude amplification algorithm solves the optimization problem to determine the minimum number of iterations $L$ and the corresponding optimal angles $\{ \varphi_n \}$ that yield a final failure probability below the desired tolerance.

\begin{figure*}[t]
\includegraphics[width=1.0\textwidth]{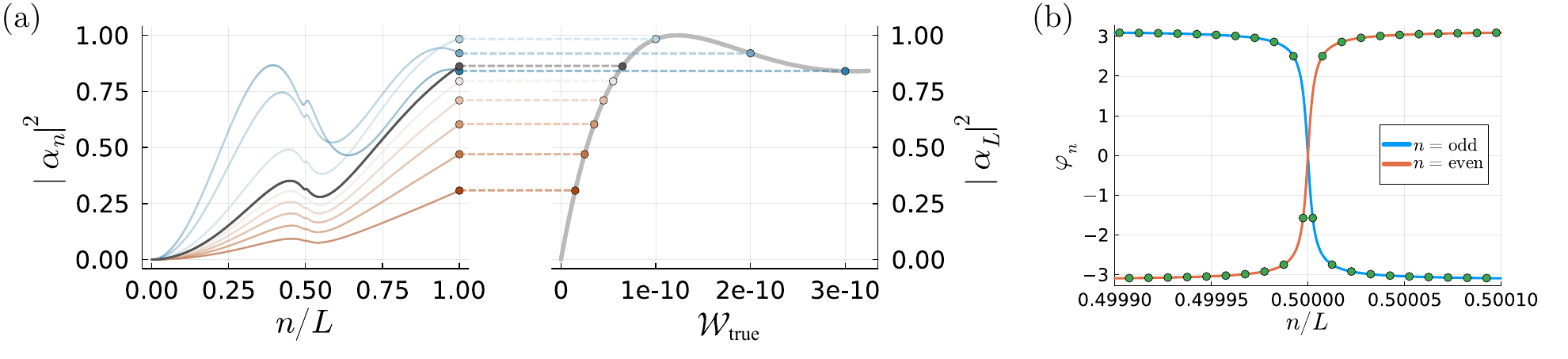}
\caption{{\bf Demonstration of the amplitude amplification algorithm.} 
(a) Evolution of the success probability $|\alpha_n|^2$ as a function of $n/L$ with $L=200001$, starting from the initial value, ${\cal W}=6.5\times 10^{-11}$, to the targeting value, $1-\delta^2$ with $\delta=0.4$, which is illustrated by the dark curve.  
In the actual implementation of the amplitude amplification algorithm, ${\cal W}$ is not known precisely beforehand and must be estimated. Consequently, its true value, ${\cal W}_{\rm true}$, might differ from the estimated one.
The light red and blue curves illustrate how $|\alpha_n|^2$ would evolve if ${\cal W}_{\rm true}$ were below or above the estimated value, ${\cal W}=6.5\times 10^{-11}$, respectively.
On the right side of the figure, the final success probability $|\alpha_{L}|^2$ is plotted as a function of ${\cal W}_{\rm true}$. 
As one can see, the final success probability is confined within the window of $(1-\delta^2,1)$, while exhibiting some oscillations in the middle of the iteration, so long as ${\cal W}_{\rm true}$ is above ${\cal W}$.
(b) Conditional rotation angles $\varphi_n$ as a function of $n/L$. 
For most of the iteration steps, the conditional rotation angles are approximately $\pi$, except when $n \simeq L/2$.}
\label{fig:Success_probability}
\end{figure*}

It is possible to obtain a complete analytic solution to this optimization problem based on a groundbreaking work~\cite{Yoder14}.
First, let us choose $\delta \in (0,1)$, which sets the desired tolerance.
Then, the optimal angles $\{ \varphi_n \}_{n=1, \ldots, L}$ are given by the following expression~\cite{Yoder14,Gilyen19,MARTYN2021}:
\begin{equation}
\varphi_n = (-1)^{n-1} \, 2 \cot^{-1} \left( \sqrt{1-\gamma^2} \tan{\frac{n \pi}{L}} \right),
\end{equation}
where the value of $\gamma$ will be determined through the optimization condition described below.
With this choice of $\{\varphi_n\}$, $\beta_L$ is in turn given as follows:
\begin{align}
\beta_L = \exp{\left(i \sum_{n=1}^{(L-1)/2} \varphi_{2n} \right)} \frac{T_L \left(\sqrt{1-{\cal W}}/\gamma \right)}{T_L \left(1/\gamma \right)},
\label{eq:Recursion-solution}
\end{align}
where $T_L$ is the $L$-th Chebyshev polynomial of the first kind, and $L$ is a specific odd integer that must be determined along with $\gamma$ to ensure that the final failure probability falls below the desired tolerance.
See Appendix~\ref{appen:Recursion_solution} for the derivation of Eq.~\eqref{eq:Recursion-solution}. 
Also, see a recent paper that provides a revised proof of the recursion relation~\cite{li2026revisiting}.

To determine $L$ and $\gamma$, let us compute the final failure probability:
\begin{equation}
|\beta_L|^2 = \delta^2 \left| T_L \left(\sqrt{1-{\cal W}}/\gamma \right) \right|^2,
\end{equation}
where we have used the condition $T_L(1/\gamma)= 1/\delta$.
Since the Chebyshev polynomial satisfies $|T_L (x)| \ge 1$ if $|x| \ge 1$ and $|T_L (x)| \le 1$ if $|x| \le 1$ for a real number $x$, we can bound the final failure probability to be $|\beta_L|^2 \le \delta^2$ if $\sqrt{1-{\cal W}}/\gamma \le 1$ holds. 
This condition can be satisfied if we choose an odd integer $L$ satisfying 
\begin{equation}
L \ge \frac{\ln(2/\delta)}{\sqrt{{\cal W}}}
\label{eq:L-condition}
\end{equation}
for small $\delta$ and large $L$~\cite{Yoder14}. 
See Appendix~\ref{appen:L-condition} for the derivation of Eq.~\eqref{eq:L-condition}.

To demonstrate how the AAGP algorithm works, we present in Fig.~\ref{fig:Success_probability} the evolution of the success probability $|\alpha_n|^2$ for the case of $(\delta, {\cal W})=(0.4, 6.5\times 10^{-11})$, which corresponds to the condition $L \ge 200001$ according to Eq.~\eqref{eq:L-condition}.
As one can see,  $|\alpha_n|^2$ evolves smoothly to the final success probability $|\alpha_L|^2$, which falls within the window of $(1-\delta^2,1)$, while exhibiting some oscillations in the middle of the iteration.

In summary, our AAGP algorithm requires ${\cal O}(1/\sqrt{\cal W})$ queries of ${\cal U}_{\rm BCS}$ and ${\cal U}^{\dagger}_{\rm BCS}$ accompanied by some specific conditional unitary operators, achieving the same quadratic speed-up as Grover's quantum search algorithm, compared to the measurement-based projection algorithm outlined in Eq.~\eqref{eq:Naive_projection}.
Note that the measurement-based projection algorithm requires ${\cal O}(1/{\cal W})$ queries of ${\cal U}_{\rm BCS}$ and the subsequent measurement of double occupancy.


\subsection{Quantum circuits for the conditional unitary operators}

In this subsection, we explain how to construct quantum circuits for the conditional unitary operators, $\mathcal{R}_{\rm vac}(\varphi)$ and $\mathcal{R}_{\rm RVB}(\varphi)$.

First, implementing $\mathcal{R}_{\rm vac}(\varphi)$ is rather straightforward using the C$^{N}$NOT gate. 
Specifically, a quantum circuit for $\mathcal{R}_{\rm vac} (\varphi)$ can be constructed using one ancilla qubit as follows:
\begin{widetext}
\begin{equation}
\begin{quantikz}[row sep={0.3cm,between origins}, column sep=0.2cm]
\lstick{\ket{0}} & \gate{X}  & \targ{} & \gate{\hat{R}_{\ket{0}}(\varphi)} & \targ{} & \gate{X} & \qw \rstick{\ket{0}} \\ [0.5cm]
\lstick[wires=6]{$\alpha \ket{{\bf 0}} + \beta \ket{{\bf 0}^\perp}$} & \gate[6]{X^{\otimes N}} & \control{} & \qw & \control{} & \gate[6]{X^{\otimes N}} & \qw 
\rstick[6]{$\alpha e^{i\varphi} \ket{{\bf 0}} + \beta \ket{{\bf 0}^\perp}$} \\
& & \control{} & \qw & \control{} & & \qw  \\
& & \control{} & \qw & \control{} & & \qw  \\
& & \control{} & \qw & \control{} & & \qw  \\
& & \control{} & \qw & \control{} & & \qw \\
& & \ctrl{-6} & \qw & \ctrl{-6} & & \qw
\end{quantikz},
\end{equation}
\end{widetext}
where $\hat{R}_{\ket{0}}(\varphi)$ is the phase shift gate that applies a phase shift of $e^{i\varphi}$ to the ancilla state $\ket{0}$.

Second, implementing ${\cal R}_{\rm RVB}(\varphi)$ directly is quite difficult as it requires identifying the RVB state.
The entire purpose of the Gutzwiller projection is to identify the RVB state within the BCS state. 
Fortunately, we can overcome this difficulty by applying the conditional unitary operator for phase rotation on the Gutzwiller-projected state, $\mathcal{R}_{\rm G}(\varphi)$, instead of $\mathcal{R}_{\rm RVB}(\varphi)$:
\begin{equation}
\mathcal{R}_{\rm G} (\varphi) = \openone + (e^{i\varphi} - 1) \mathcal{P}_{\rm G},
\end{equation}
where $\mathcal{P}_{\rm G}$ is the Gutzwiller projection operator that removes all states with double occupancy.
It is important to note that $\mathcal{R}_{\rm G}(\varphi)$ is identical to $\mathcal{R}_{\rm RVB}(\varphi)$ on the effective two-dimensional Hilbert space spanned by $\vert \psi_{\rm RVB} \rangle$ and $\vert \psi_{\rm RVB}^\perp \rangle$.

Specifically, a quantum circuit for $\mathcal{R}_{\rm G}(\varphi)$ can be constructed using $N_s+1$ ancilla qubits as follows:
\begin{widetext}
\begin{equation}
\begin{quantikz}[row sep={0.3cm,between origins}, column sep=0.2cm]
\lstick{\ket{0}} & \qw & \qw & \qw & \gate{X} & \targ{} & \gate{\hat{R}_{\ket{0}} (\varphi)} & \targ{} & \gate{X} & \qw & \qw & \qw & \qw \rstick{\ket{0}} \\ [0.4cm]
\lstick[wires=3]{$\ket{0}^{\otimes N_s}$} & \targ{} & \qw & \qw & \gate[3]{X^{\otimes N_s}} & \control{} & \qw & \control{} & \gate[3]{X^{\otimes N_s}} & \qw & \qw & \targ{} & \qw \rstick[wires=3]{$\ket{0}^{\otimes N_s}$} \\ 
& \qw & \targ{} & \qw & & \control{} & \qw & \control{} & & \qw & \targ{} & \qw & \qw \\
& \qw & \qw & \targ{} & & \ctrl{-3} & \qw & \ctrl{-3} & & \targ{} & \qw & \qw & \qw \\ [0.4cm]
\lstick[wires=6]{$\alpha \ket{\psi_{\rm RVB}}  + \beta \ket{\psi_{\rm RVB}^\perp}$ } & \control{} & \qw & \qw & \qw & \qw & \qw & \qw & \qw & \qw & \qw & \control{} & \qw \rstick[wires=6]{$\alpha e^{i\varphi} \ket{\psi_{\rm RVB}} + \beta \ket{\psi_{\rm RVB}^\perp}$} \\
& \ctrl{-4} & \qw & \qw & \qw & \qw & \qw &\qw & \qw & \qw & \qw & \ctrl{-4} & \qw  \\
& \qw & \control{} & \qw & \qw & \qw & \qw & \qw & \qw & \qw & \control{} & \qw & \qw \\
& \qw & \ctrl{-5} & \qw & \qw & \qw & \qw & \qw & \qw & \qw & \ctrl{-5} & \qw & \qw \\
& \qw & \qw & \control{} & \qw & \qw & \qw & \qw & \qw & \control{} & \qw & \qw & \qw \\
& \qw & \qw & \ctrl{-6} & \qw & \qw & \qw & \qw & \qw & \ctrl{-6} & \qw & \qw & \qw 
\end{quantikz},
\end{equation}
\end{widetext}
where it is interesting to note that the upper central part of the circuit, which contains $N_s + 1$ ancilla qubits, has a structure similar to $\mathcal{R}_{\rm vac} (\varphi)$, except that the number of qubits here is $N_s+1$ instead of $N+1$.
Once again, $N=2 N_s$ due to the spin degree of freedom.

\subsection{Quantum resource requirements}

Since many different quantum gates are used in our quantum circuits, it is important to estimate the total quantum resource requirements.
To this end, we estimate the number of $T$-gates as a measure of the non-Clifford gates required to implement our quantum circuits.
Two main non-Clifford gates used in our circuit are single-qubit Pauli rotations and C$^N$NOT gates.

In the current state-of-the-art algorithm~\cite{Ross16}, a typical single-qubit Pauli $z$-rotation requires $3 \log_2(1/\epsilon)$ $T$-gates, where $\epsilon$ determines the precision. 
As our circuits use many single-qubit Pauli rotations, we assume that each is typical in terms of $T$-gate count.
The C$^N$NOT gate can be implemented using $2N$ Toffoli gates and one CNOT gate with $N$ ancilla qubits~\cite{Nielsen_Chuang_Book}. 
Given that the optimal count of $T$-gates for the Toffoli gate is $7$~\cite{Amy13, Gosset14}, the $T$-gate count for the C$^N$NOT gate scales as ${\cal O}(N)$.

Now, each Givens rotation involves two single Pauli rotations. 
This means that the quantum circuit for $\mathcal{U}_{\rm BCS}$ has a $T$-gate count of ${\cal O}(N^2 \log_2 (1/\epsilon))$ when approximating each Pauli rotation to a precision of $\epsilon$.
Note that, as shown previously, the number of two-qubit gates in $\mathcal{U}_{\rm BCS}$ is $(N-n)(n+2p) - 2p^2$, which scales as ${\cal O}(N^2)$ since both $n$ and $p$ scale as ${\cal O}(N)$.

Meanwhile, the quantum circuit for $\mathcal{R}_{\rm vac} (\varphi)$ uses two C$^N$NOT gates and one Paili rotation gate, resulting in a $T$-gate count of ${\cal O}(N) + {\cal O}(\log_2(1/\epsilon))$.
Similarly, the quantum circuit for $\mathcal{R}_{\rm G} (\varphi)$ uses $N_s$ Toffoli gates, two C$^{N_s}$NOT gates, and one Pauli rotation gate, resulting in a similar $T$-gate count of ${\cal O}(N) + {\cal O}(\log_2(1/\epsilon))$.
For large $N$, the $T$-gate counts for $\mathcal{R}_{\rm vac} (\varphi)$ and $\mathcal{R}_{\rm G} (\varphi)$ are negligible compared to those for $\mathcal{U}_{\rm BCS}$ and $\mathcal{U}^\dagger_{\rm BCS}$.

Finally, the quantum circuit for the AAGP algorithm requires ${\cal O}\left(\frac{\ln (2/\delta)}{\sqrt{\mathcal{W}}}\right)$ queries of $\mathcal{U}_{\rm BCS}$ and $\mathcal{R}_{\rm G} (\varphi)$, or $\mathcal{U}^\dagger_{\rm BCS}$ and $\mathcal{R}_{\rm vac} (\varphi)$.
This means that the total $T$-gate count of the quantum circuit for the RVB state is ${\cal O} \left(\frac{\ln (2/\delta)}{\sqrt{\cal W}} N_s^2 \log_2 (1/\epsilon)  \right)$.

\section{Probability weight of the RVB state within the BCS state}
\label{sec:RVB_weight}

The probability weight of the RVB state within the BCS state is given by the square of the overlap between the two states.
\begin{align}
{\cal W}=|\langle \psi_{\rm RVB}|\psi_{\rm BCS} \rangle |^2 ,
\label{eq:W_def}
\end{align}
where $\psi_{\rm RVB}$ and $\psi_{\rm BCS}$ denote the RVB and BCS wave functions, respectively.  
The wave function of the RVB state is related to that of the BCS state via the Gutzwiller projection:
\begin{align}
|\psi_{\rm RVB}\rangle = {\cal C} {\cal P}_{\rm G} | \psi_{\rm BCS}\rangle ,
\end{align}
where 
${\cal C}$ is the normalization constant.
We can calculate ${\cal W}$ based on its relationship with ${\cal C}$:
\begin{align}
{\cal W}=1/|{\cal C}|^2 .
\label{eq:W_vs_C}
\end{align}
To understand why this is true, let us rewrite ${\cal W}$ in Eq.~\eqref{eq:W_def} as follows:
\begin{align}
{\cal W} &= |{\cal C}|^2 |\langle \psi_{\rm BCS}|{\cal P}_{\rm G}|\psi_{\rm BCS} \rangle |^2 
\nonumber \\
&= |{\cal C}|^2 |\langle \psi_{\rm BCS}|{\cal P}_{\rm G}^2|\psi_{\rm BCS} \rangle |^2 ,
\label{eq:W_derivation1}
\end{align}
where we have used the fact that ${\cal P}_G^2={\cal P}_G$.
Then, by converting the BCS wave function back to the RVB wave function, one can rewrite Eq.~\eqref{eq:W_derivation1} as follows:  
\begin{align}
{\cal W} &=\frac{1}{|{\cal C}|^2} |\langle \psi_{\rm RVB}|\psi_{\rm RVB} \rangle |^2 ,
\label{eq:W_derivation2}
\end{align}
which reduces to Eq.~\eqref{eq:W_vs_C} due to the normalization of the RVB wave function.

The next step is to determine how to calculate ${\cal C}$.
To this end, let us examine exactly how ${\cal P}_G$ operate on the BCS wave function, which can be expanded as follows:
\begin{align}
|\psi_{\rm BCS}\rangle &=  \sum_{n \in \textrm{GPHS}} a_n |\phi_{n}\rangle +\sum_{n \notin \textrm{GPHS}} a_n |\phi_n\rangle ,
\label{eq:BCS_expanded}
\end{align}
where $\{|\phi_n\rangle\}$ is, at the moment, any convenient set of basis states, categorized into two groups: one group that is included in the Gutzwiller-projected Hilbert space (GPHS) and the other that is not. 
In other words, the first and second groups on the right side of Eq.~\eqref{eq:BCS_expanded} represent the components of the BCS wave function that survive and vanish after the application of ${\cal P}_{\rm G}$, respectively.
In this representation, the RVB wave function can be represented by
\begin{align}
|\psi_{\rm RVB}\rangle &=  {\cal C} \sum_{n \in \textrm{GPHS}} a_n |\phi_{n}\rangle ,
\label{eq:RVB_expanded1}
\end{align}
where ${\cal C}$ is the normalization constant that we aim to calculate.

It is now convenient to represent the BCS wave function using a set of basis states that distinguish the vacuum state from others.
In this set of basis states, the amplitude of the vacuum state in the RVB wave function is given by ${\cal C} a_0$, where $a_0$ refers to the amplitude of the vacuum state in the BCS wave function.
As shown below, ${\cal C}$ can be calculated by taking the ratio of the amplitudes of the vacuum state in the RVB and BCS wave functions.

Suppose that we have the explicit form of the RVB wave function as follows:
\begin{align}
|\psi_{\rm RVB}\rangle &=  \sum_{n \in \textrm{GPHS}} b_n |\phi_{n}\rangle  ,
\label{eq:RVB_expanded2}
\end{align}
where $\{b_n\}$ are the amplitudes of the RVB wave function in the same set of basis states used to represent the BCS wave function, $\{|\phi_n\rangle\}$.

There is a particularly convenient set of basis states that can be formed by considering all possible arrangements of spin-up and spin-down electrons in a given lattice, subject to the no-double-occupancy constraint imposed by the Gutzwiller projection.
Specifically, let us define $|\phi_n\rangle$ as the $n$-th configuration of electrons, where spin-up and spin-down electrons are distributed in the lattice sites at $\{ {\bf r}^{(n)}_{\uparrow,i} \}$ and $\{ {\bf r}^{(n)}_{\downarrow,j} \}$, respectively, with $i$ and $j$ enumerating the lattice sites occupied by the corresponding spin species. 
Note that $\{ {\bf r}^{(n)}_{\uparrow,i} \}$ and $\{ {\bf r}^{(n)}_{\downarrow,j} \}$ are mutually exclusive due to the no-double-occupancy constraint.
It is important to note that the total number of electrons can range from zero to the total number of lattice sites, covering all possible configurations from vacuum to half-filling.

It has been previously shown~\cite{Kwon22, Gros88} that, in this basis, the amplitudes of the RVB wave function can be expressed as follows, up to an overall normalization constant:
\begin{align}
b_n \propto {\rm det}[\varphi^{(n)}_{ij}] ,
\end{align}
where $\varphi^{(n)}_{ij}=\varphi({\bf r}^{(n)}_{\uparrow,i}-{\bf r}^{(n)}_{\downarrow,j})$ is the real-space component of the Cooper pair wave function with the constituent spin-up and spin-down electrons located at ${\bf r}^{(n)}_{\uparrow,i}$ and ${\bf r}^{(n)}_{\downarrow,j}$ in the $n$-th configuration, respectively.
Note that $\varphi({\bf r})$ can be computed by performing the Fourier transform of $v_{\bf k}/u_{\bf k}$, 
where $u_{\bf k}$ and $v_{\bf k}$ are the usual parameters of the BCS wave function,
\begin{align}
u^2_{\bf k} =\frac{1}{2} \left(
1+\frac{\xi_{\bf k}}{\sqrt{\xi^2_{\bf k}+\Delta^2_{\bf k}}}
\right)
\end{align}
and
\begin{align}
v^2_{\bf k} =\frac{1}{2} \left(
1-\frac{\xi_{\bf k}}{\sqrt{\xi^2_{\bf k}+\Delta^2_{\bf k}}}
\right)
\end{align}
with $\xi_{\bf k}=\epsilon_{\bf k}-\mu$ being the difference between the kinetic energy and the chemical potential energy, and $\Delta_{\bf k}$ being the pairing amplitude.
Here, we use the simple nearest-neighbor hopping model with $\epsilon_{\bf k}=-2t(\cos{k_x}+\cos{k_y})$, where $t$ is the hopping parameter.
Also, we focus on $d$-wave superconductivity by setting $\Delta_{\bf k}=2\Delta(\cos{k_x}-\cos{k_y})$, where $\Delta$ is the variational gap parameter.

\begin{figure*}[t]
\includegraphics[width=0.9\textwidth]{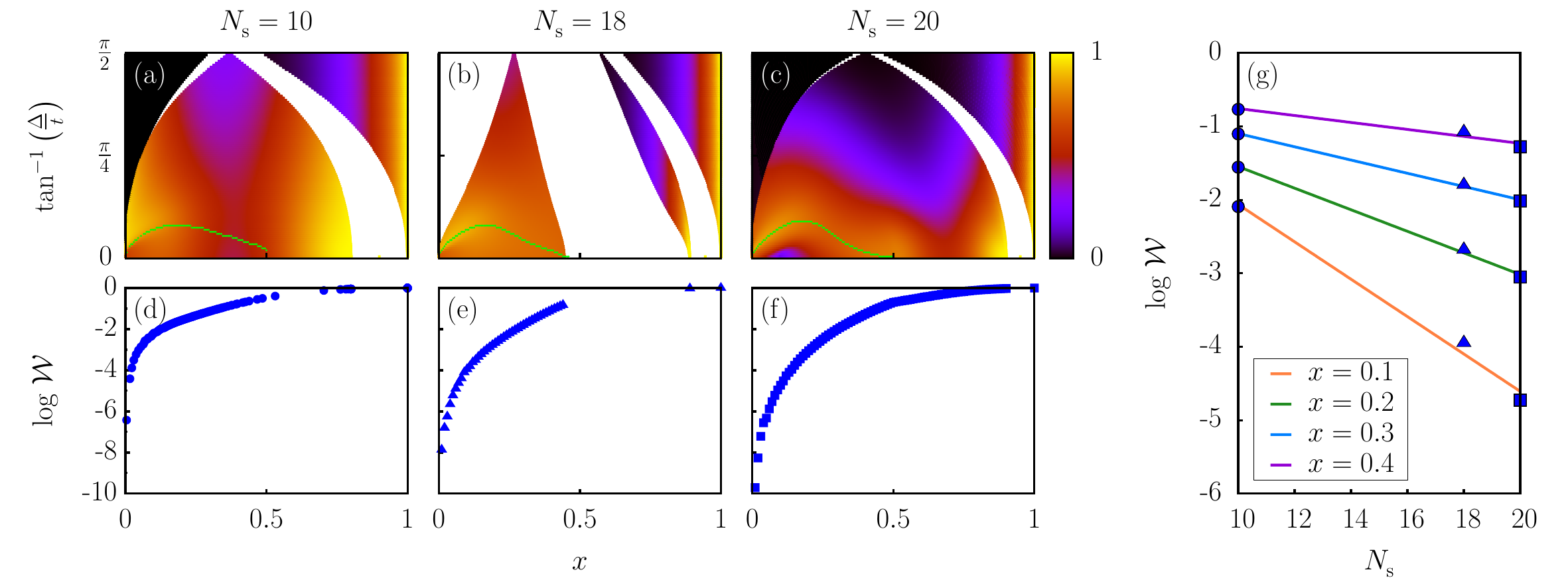}
\caption{{\bf Probability weight of the RVB state within the BCS state.} 
Panels ({\bf a}), ({\bf b}), and ({\bf c}) show the phase diagram of the overlap between the RVB state and the exact ground state of the $t$-$J$ model as a function of pairing amplitude $\Delta$ and hole doping concentration $x$ for finite-size systems with $N_s=10$, $18$, and $20$, respectively.
Note that $N_s$ represents the number of sites in finite-size systems.
To cover the full range of pairing amplitude from zero to infinity, the $y$-axis of Panels ({\bf a}), ({\bf b}), and ({\bf c}) is scaled as $\tan^{-1}(\Delta/t)$.
The color bar on the right shows the overlap scale, with bright yellow indicating unity and black indicating zero.
Green curves trace the maximum overlap as a function of $x$ for each $N_s$.
Panels ({\bf d}), ({\bf e}), and ({\bf f}) show the logarithm of the probability weight of the RVB state within the BCS state, $\log{\cal W}$, as a function of $x$ along the maximum overlap curves for $N_s=10$, $18$, and $20$, respectively.
Note that, for $x \geq 0.5$, the maximum overlap occurs along the line of zero $\Delta$.
Panel ({\bf g}) shows the linear scaling of $\log{\cal W}$ as a function of $N_s$ at $x=0.1$, 0.2, 0.3, and 0.4.
}
\label{fig:RVB_weight}
\end{figure*}

Accounting for the overall normalization constant, the amplitudes of the $n$-the configuration in the RVB wave function can be expressed as follows:
\begin{align}
b_n = {\cal N} {\rm det}[\varphi^{(n)}_{ij}] ,
\end{align}
where the normalization constant ${\cal N}$ is given by
\begin{align}
1/{\cal N}= \sqrt{ \sum_{n \in {\rm GPHS}} \left|{\rm det}[\varphi^{(n)}_{ij}]\right|^2} .
\end{align}
This shows that the amplitude of the vacuum state in the RVB wave function is given by $b_0$, which is equivalent to ${\cal N}$ because the determinant part of the vacuum state is unity. 
In other words, we have $b_0 = {\cal N}$.

By comparing Eqs.~\eqref{eq:RVB_expanded1} and \eqref{eq:RVB_expanded2}, it becomes evident that the amplitude of the vacuum state in the RVB wave function can be expressed in two distinct forms, which should ultimately be identical, so that
\begin{align}
{\cal C} a_0=b_0 ,
\end{align}
or equivalently
\begin{align} 
{\cal W}=1/|{\cal C}|^2=|a_0/b_0|^2=|a_0/{\cal N}|^2 ,
\end{align}
indicating that ${\cal W}$ can be calculated once $a_0$ and ${\cal N}$ are known.

Now, it is straightforward to determine $a_0$ from the standard form of the BCS wave function:
\begin{align}
|\psi_{\rm BCS}\rangle = \prod_{\bf k} (u_{\bf k}+v_{\bf k} c^\dagger_{{\bf k},\uparrow} c^\dagger_{-{\bf k},\downarrow}) |0\rangle , 
\end{align} 
where $|0\rangle$ is the vacuum state.
The amplitude of the vacuum state is then given by
\begin{align}
a_0=\prod_{\bf k} u_{\bf k} .
\end{align}
Combining everything together, the probability weight of the RVB state within the BCS state is finally given by
\begin{align}
{\cal W}=\prod_{\bf k} u^2_{\bf k} \left( \sum_{n \in {\rm GPHS}} \left|{\rm det}[\varphi^{(n)}_{ij}]\right|^2 \right),
\end{align}
which is uniquely determined by the variational gap parameter $\Delta$ and the chemical potential $\mu$.

We aim to calculate ${\cal W}$ when the RVB state is relevant to high-temperature superconductivity.  We consider the uniform $t$-$J$ model on a square lattice which can be derived from the infinite repulsion limit of an on-site Hubbard model \cite{auerbach1994interacting}.  The $t$-$J$ model is expected to harbor critical features observed in high-temperature superconductors and may also contain the fundamental mechanism underlying pairing from repulsion. The $t$-$J$ model is also more efficient than the Hubbard model to implement in quantum algorithms \cite{MYERS2023}.  

To find ${\cal W}$ we must accurately construct the BCS input parameters.  Specifically, we must find $\Delta$ as a function of hole doping concentration, $x$, that maximizes the overlap between the RVB state and the exact ground state of the $t$-$J$ model at each $x$.  Note that we initially calculate the overlap between the RVB state and the exact ground state of the $t$-$J$ model at specified values of $\Delta$ and $\mu$, and then convert it to a function of $\Delta$ and $x$.

The overlap between the RVB state and the exact ground state of the $t$-$J$ model can be computed using straightforward exact diagonalization methods, such as the modified Lanczos method.
However, there is a subtlety in properly defining the overlap due to the inherent fluctuations in particle number in the RVB state. 
Specifically, the RVB state can be expressed as a sum of its particle-number components as follows:
\begin{equation}
| \psi_{\rm RVB} \rangle = \sum_N \lambda_N | \phi^{N}_{\rm RVB} \rangle  ,
\end{equation}
where $\lambda_N$ and $| \phi^{N}_{\rm RVB} \rangle$ denote the amplitude and the component of the RVB state in the specific particle-number sector with $N$ particles, respectively.
 On the other hand, the exact ground state of the $t$-$J$ model exhibits no such fluctuations.
To account for this difference, we define proper overlap as the average of the square of the overlap between the two states in each particle-number sector $N$, weighted by the probability weight $|\lambda_N|^2$~\cite{Kwon22}:
\begin{equation}
\overline{{\cal O}^2}=\sum_N |\lambda_N|^2 |\langle \phi^{N}_{\rm RVB} | \phi^{N}_{t\mbox{-}J} \rangle|^2 ,
\end{equation}
where $| \phi^{N}_{t\mbox{-}J} \rangle$ is the exact ground state of the $t$-$J$ model in the sector with $N$ particles.
 For simplicity, we will refer to $\overline{{\cal O}^2}$ as simply the overlap from now on.

Figure~\ref{fig:RVB_weight} presents the results of numerical analyses to determine ${\cal W}$ across various finite-size systems up to the maximum obtainable within classical memory constraints.
Specifically, panels ({\bf a}), ({\bf b}), and ({\bf c}) in Fig.~\ref{fig:RVB_weight} show the phase diagram of the overlap as a function of $\Delta$ and $x$ for finite-size systems with the number of states, $N_s=10$, $18$, and $20$, respectively.
We calculate ${\cal W}$ along the green curves, where the overlap is maximized at each $x$.
Panels ({\bf d}), ({\bf e}), and ({\bf f}) in Fig.~\ref{fig:RVB_weight} show $\log{\cal W}$ as a function of $x$ along the maximum-overlap curves for $N_s=10$, $18$, and $20$, respectively.
Finally, panel ({\bf g}) shows that $\log{\cal W}$ scales linearly with $N_s$ for various values of $x$.
Consequently, ${\cal W}$ can be expressed as follows:
\begin{equation}
{\cal W} = {\cal A} e^{-\lambda N_s},
\end{equation}
where ${\cal A}$ and $\lambda$ denote the exponential prefactor and the decay constant of ${\cal W}$, which depends on the hole doping concentration $x$.
For $x=0.1$, $0.2$, $0.3$, and $0.4$, ${\cal A}$ is $3.0452$, $0.8359$, $0.6373$, and $0.5228$, and $\lambda$ is $0.5861$, $0.3382$, $0.2075$, and $0.1092$, respectively.

This means that, for example, at $N_s=100$, ${\cal W} \simeq 1.7 \times 10^{-15}$ for $x=0.2$, which yields the largest pairing gap and is therefore considered as optimal doping within the RVB framework.
Note that $N_s=100$ is sufficiently large to exhibit near-thermodynamic behavior.
Using the measurement-based projection algorithm, this would require checking for double occupancy at every site an order of $10^{15}$ times to achieve the Gutzwiller projection. 
On the other hand, using the AAGP algorithm developed in this work, the Gutzwiller projection can be achieved by applying ${\cal U}_{\rm BCS}$ and ${\cal U}^{-1}_{\rm BCS}$ (with each accompanied by an appropriate conditional unitary operator) only an order of $10^{7}$ or $10^{8}$ times.


\section{SUMMARY AND OUTLOOK}
\label{sec:discussions}

In this work, we have constructed scalable quantum algorithms for implementing the Gutzwiller projection via amplitude amplification.
The AAGP algorithm constructed here reduces the number of qubit operations by many orders of magnitude compared with the measurement-based projection algorithm, thereby making it a potentially useful input-state preparation primitive for quantum algorithms targeting strongly correlated models. The key point is not only the asymptotic quadratic reduction from ${\cal O}(1/{\cal W})$ to ${\cal O}(1/\sqrt{\cal W})$, but also that, in the regime benchmarked here, this reduction is large enough to cross a practical finite-size threshold. For projected BCS/RVB states relevant to our test case, the improvement changes the state-preparation step from effectively unusable to plausibly deployable at lattice sizes of physical interest.

We emphasize, however, that AAGP by itself does not resolve the full resource challenge of simulating strongly correlated models. Rather, it removes a key obstruction in preparing a physically motivated class of input states. This distinction is important because the usefulness of many quantum simulation algorithms depends sensitively on the choice of the initial state, and projected BCS/RVB states provide a flexible ansatz family for strongly correlated Hubbard, $t$-$J$, and related spin models. In this sense, AAGP should be viewed as an enabling primitive within a broader fault-tolerant workflow. 

The finite-size regime emphasized here is already nontrivial. The scaling analysis suggests that the input RVB state remains meaningful when extrapolated to lattice sizes on the order of 100 sites, where direct classical overlap calculations are no longer feasible. Thus, although the exponentially small RVB weight continues to govern the asymptotic cost, the quadratic improvement is sufficient to make an important difference at the physically relevant finite sizes that motivate quantum simulation in the first place.

The initial states discussed here strictly enforce Gutzwiller projection, but this can be relaxed. No double occupancy is precisely the limit assumed by the $t$-$J$ model, where the system is simplified by taking the on-site repulsive interaction, $U$, to be infinitely large.  A natural question is whether it is possible to build a similar ansatz for the ground states of Hubbard models that allow the controlled addition of double occupancies. One possible extension is to use AAGP and then perform a systematic unitary transformation on the Gutzwiller-projected BCS state to perturbatively account for the effects of double occupancy within the $t/U$ expansion of the Hubbard model. Specifically, we can modify the Gutzwiller-projected BCS state to incorporate the effects of double occupancy perturbatively~\cite{Paramekanti01}:
\begin{equation}
|\psi\rangle = e^{iS} {\cal P}_{\rm G} |\psi_{\rm BCS}\rangle,
\end{equation}
where the unitary transformation $e^{iS}$ can be derived systematically via the $t/U$ expansion of the Hubbard model~\cite{Gros87,MacDonald88}.
Developing efficient quantum algorithms for this unitary transformation would be valuable in future research.

\begin{acknowledgments}

The authors are grateful to Yunkyu Bang and Soonwon Choi for various insightful discussions.
The authors also thank the Center for Advanced Computation (CAC) at Korea Institute for Advanced Study (KIAS) for providing computing resources for this work.
This work is partially supported by the KIAS Individual Grant, PG032303 (K.P.). 
Also, V.W.S. and K.P. acknowledge support from the joint grant under the US-Korea Quantum Initiative by the US Air Force Office of Scientific Research (AFOSR), FA2386-21-1-4081 and the National Research Foundation of Korea (NRF) funded by the Ministry of Science and ICT of the Korean Government, NRF-2021M3H3A1085208.  The contributions of V.W.S. 
to this work were supported in part by the U.S. Department of Energy, Office of Science, Office of Basic Energy Sciences Energy Frontier Research Centers program under Award Number DE-SC-0026289 and in part by AFOSR (FA9550-19-1-0272 and FA9550-23-1-0034) and ARO (W911NF2210247).  

\end{acknowledgments}

\appendix

\section{Redundancy in the Bloch-Messiah decomposition}
\label{appen:Redundancy}

The Bloch-Messiah decomposition contains redundancy, so that different forms of $\mathbf{D}|_{n+2p}$ can yield the same BCS state.
Specifically, the BCS state in Eq.~\eqref{eq:BCS} is invariant under the following $\mathrm{SU}(n)$ transformation acting on unpaired states: 
\begin{align}
\tilde{f}_i^\dagger = 
\begin{cases} 
\sum\limits_{j=1}^n [\mathbf{M}]_{ji} f_j^\dagger \qquad \textrm{if } 1 \le i \le n, \\
f_i^\dagger \qquad \qquad \qquad \textrm{if } i > n,
\end{cases}
\end{align}
where $\mathbf{M} \in \mathrm{SU}(n)$.
This invariance is due to the fact that $\tilde{f}_1^\dagger \ldots \tilde{f}_n^\dagger = f_1^\dagger \ldots f_n^\dagger$. 
In matrix form, the above $\mathrm{SU}(n)$ transformation acts on $\mathbf{D}|_{n+2p}$ as follows:
\begin{equation}
\mathbf{D}|_{n+2p} \mapsto \mathbf{D}|_{n+2p} 
\left( 
\begin{array}{c|c}
\mathbf{M} & \\ \cline{1-2}
 & \mathbf{1}_{2p \times 2p}
\end{array} 
\right),
\end{equation}
which can be in turn used to make $[\mathbf{D}|_{n+2p}]_{ij}=0$ for $i - j > N - n$. 

For example, when $N=6$, $n=3$, and $p=1$, $\mathbf{M}$ can be chosen such that

\begin{widetext} 
\begin{align}
\mathbf{D} = \left( \begin{array}{ccc|cc}
* & * & * & * & * \\
* & * & * & * & * \\
* & * & * & * & * \\
* & * & * & * & * \\
* & * & * & * & * \\
* & * & * & * & * 
\end{array} \right) &\mapsto \left( \begin{array}{ccc|cc}
* & * & * & * & * \\
* & * & * & * & * \\
* & * & * & * & * \\
* & * & * & * & * \\ \cline{1-2}
\lvertline{*} & \rvertline{*} & * & * & * \\
\lvertline{*} & \rvertline{*} & * & * & * \\ \cline{1-2}
\end{array} \right) \left( \begin{array}{ccc|cc}
\cline{1-2}
\lvertline{e^{i\phi_a} \cos \theta_a} & \rvertline{e^{i \phi_a} \sin \theta_a} & 0 & & \\
\lvertline{-e^{-i \phi_a} \sin \theta_a} & \rvertline{e^{-i \phi_a} \cos \theta_a} & 0 & & \\ \cline{1-2}
0 & 0 & 1 & & \\
\hline
& & & 1 & 0 \\
& & & 0 & 1
\end{array}
\right) = \left( \begin{array}{ccc|cc}
* & * & * & * & * \\
* & * & * & * & * \\
* & * & * & * & * \\
* & * & * & * & * \\ \cline{1-2}
\lvertline{*} & \rvertline{*} & * & * & * \\
\lvertline{{\bf 0}} & \rvertline{*} & * & * & * \\ \cline{1-2}
\end{array} \right) \nonumber \\
&\mapsto \left( \begin{array}{ccc|cc}
* & * & * & * & * \\
* & * & * & * & * \\
* & * & * & * & * \\
* & * & * & * & * \\ \cline{2-3}
* & \lvertline{*} & \rvertline{*} & * & * \\
0 & \lvertline{*} & \rvertline{*} & * & * \\ \cline{2-3}
\end{array} \right) \left( \begin{array}{ccc|cc}
1 & 0 & 0 & & \\ \cline{2-3}
0 & \lvertline{e^{i\phi_b} \cos \theta_b} & \rvertline{e^{i \phi_b} \sin \theta_b} & & \\
0 & \lvertline{-e^{-i \phi_b} \sin \theta_b} & \rvertline{e^{-i \phi_b} \cos \theta_b} & & \\
\hline
& & & 1 & 0 \\
& & & 0 & 1
\end{array}
\right) = \left( \begin{array}{ccc|cc}
* & * & * & * & * \\
* & * & * & * & * \\
* & * & * & * & * \\
* & * & * & * & * \\ \cline{2-3}
* & \lvertline{*} & \rvertline{*} & * & * \\
0 & \lvertline{{\bf 0}} & \rvertline{*} & * & * \\ \cline{2-3}
\end{array} \right) \nonumber \\
&\mapsto \left( \begin{array}{ccc|cc}
* & * & * & * & * \\
* & * & * & * & * \\
* & * & * & * & * \\ \cline{1-2}
\lvertline{*} & \rvertline{*} & * & * & * \\
\lvertline{*} & \rvertline{*} & * & * & * \\ \cline{1-2}
0 & 0 & * & * & * 
\end{array} \right) \left( \begin{array}{ccc|cc}
\cline{1-2}
\lvertline{e^{i\phi_c} \cos \theta_c} & \rvertline{e^{i \phi_c} \sin \theta_c} & 0 & & \\
\lvertline{-e^{-i \phi_c} \sin \theta_c} & \rvertline{e^{-i \phi_c} \cos \theta_c} & 0 & & \\ \cline{1-2}
0 & 0 & 1 & & \\
\hline
& & & 1 & 0 \\
& & & 0 & 1 
\end{array}
\right) = \left( \begin{array}{ccc|cc}
* & * & * & * & * \\
* & * & * & * & * \\
* & * & * & * & * \\ \cline{1-2}
\lvertline{*} & \rvertline{*} & * & * & * \\
\lvertline{{\bf 0}} & \rvertline{*} & * & * & * \\ \cline{1-2}
0 & 0 & * & * & * 
\end{array} \right),
\end{align}
\end{widetext}
where $*$ indicates some arbitrary numbers, and ${\bf 0}$ is the zero element vanishing due to the suitable choice of $\mathrm{SU}(n)$ rotation angles, i.e., $\theta_a$, $\theta_b$, $\theta_c$ and $\phi_a$, $\phi_b$, $\phi_c$.
As mentioned in the main text, the procedure above can be generalized to show $[\mathbf{D}|_{n+2p}]_{ij}=0$ for $i - j > N - n$.

\section{Two-qubit gate number and circuit depth of the BCS quantum circuit}
\label{appen:Two-qubit_counting}

\begin{figure*}[t]
\includegraphics[width=0.6\textwidth,angle=0]{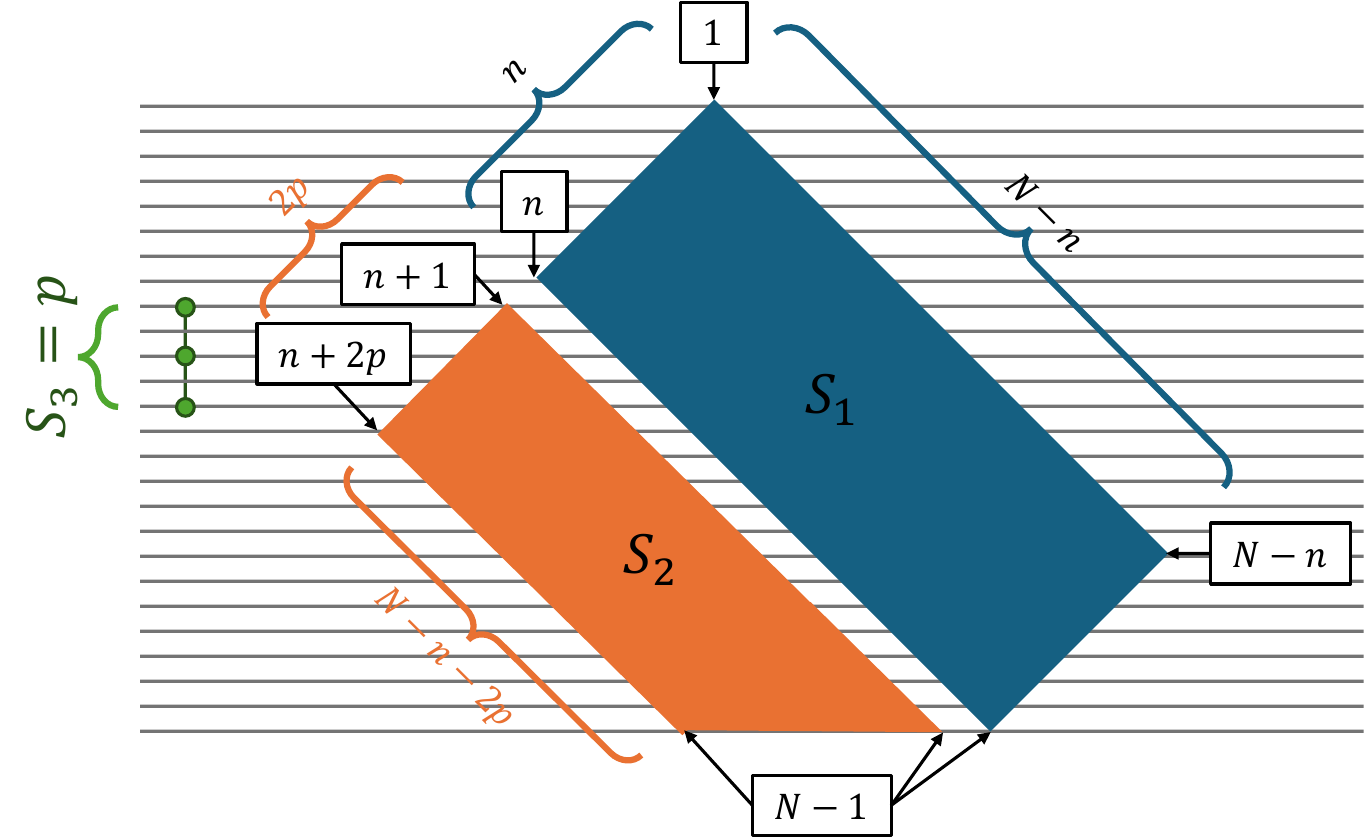}
\caption{
{\bf Schematic diagram of the Bloch-Messiah decomposition generating the full BCS state.}
Horizontal lines represent individual qubits.
Numbers inside the boxes indicate the qubit index, while numbers outside the curly brackets indicate the number of qubits within the enclosed length.
}
\label{fig:Two-qubit_counting}
\end{figure*}

In this appendix, we calculate the two-qubit gate number and the circuit depth of the BCS quantum circuit.
To begin, it is important to note that our BCS quantum circuit constructs the full BCS state using two distinct groups of two-qubit gate operations. 
The first group is associated with Givens rotations, as described in Eqs.~\eqref{eq:Givens_rotation1} and \eqref{eq:Givens_rotation2}, while the second group is associated with the creation of Cooper pairs in the depth-1 BCS state, as described in Eq.~\eqref{eq:BCS_depth-1_v2_JW}.
The first group can be further divided into two additional subgroups: Givens rotations involving (i) unpaired states and (ii) paired states.

Generalizing the example given in the main text using Eqs.~\eqref{eq:Givens_rotation1} and \eqref{eq:Givens_rotation2}, we can illustrate our BCS quantum circuit with a schematic diagram shown in Fig.~\ref{fig:Two-qubit_counting}.
As one can see from Fig.~\ref{fig:Two-qubit_counting}, the two-qubit gate number for Givens rotations involving unpaired states is $S_1=n(N-n)$, while that for Givens rotations involving paired states is 
\begin{align}
S_2 &=\sum_{k=1}^{2p} (N-n-k) 
\nonumber \\ 
&=2p(N-n)-p(2p+1).
\end{align}
In the meantime, the two-qubit gate number associated with the creation of Cooper pairs in the depth-1 BCS state is simply $S_3=p$.
When all the groups are combined, the total number of two-qubit gates is given as follows:
\begin{align}
S_{\rm tot} &= S_1+S_2+S_3
\nonumber \\
&=(N-n)(n+2p)-2p^2.
\end{align}

Finally, the circuit depth is given by the horizontal length of the quantum circuit in Fig.~\ref{fig:Two-qubit_counting}:
\begin{align}
L =N+2p.
\end{align}

\section{Solution of the amplitude amplification recursion relation}
\label{appen:Recursion_solution}

Here, we demonstrate that the recursion relation in Eq.~\eqref{eq:Beta-recursion} has the solution in Eq.~\eqref{eq:Recursion-solution}.
To this end, let us consider the following recursion relation:
\begin{equation}
a_n = x (1- e^{i \phi_{n-1}} ) a_{n-1} + e^{i \phi_{n-1}} a_{n-2},
\label{eq:Recursion1}
\end{equation}
where
\begin{equation}
\phi_n \equiv 2 \cot^{-1} \left( \sqrt{1-\gamma^2} \tan{\frac{n \pi}{L}}  \right)
\end{equation}
with $\gamma \in (0, 1)$ and $L$ being an integer.
Then, given the initial condiction $(a_0, a_1) = (1, x)$, it has been previously shown~\cite{Yoder14} that
\begin{equation}
a_L = \frac{T_L (x/\gamma)}{T_L(1/\gamma)},
\end{equation} 
where $T_L$ is the $L$-th Chebyshev polynomial of the first kind.

What we aim to demonstrate is that the recursion relation in Eq.~\eqref{eq:Recursion1} is equivalent to that in Eq.~\eqref{eq:Beta-recursion} when considering only an odd number of iteration steps.
To this end, let us rewrite Eq.~\eqref{eq:Beta-recursion} as follows:
\begin{equation}
b_n = \sqrt{1-x^2} (e^{i \varphi_{n-1}} -1) b_{n-1} + e^{i \varphi_{n-1}} b_{n-2} ,
\label{eq:Recursion2}
\end{equation}
where $\varphi_n=(-1)^{n-1}\phi_n$ and $x=\sqrt{1-{\cal W}}$.
The final goal is to prove that
\begin{equation}
b_L = \exp{\left(i \sum_{n=1}^{(L-1)/2} \varphi_{2n} \right)} \frac{T_L (x/\gamma)}{T_L(1/\gamma)},
\label{eq:Final_goal}
\end{equation} 
given the initial condition $(b_0, b_1) = (0, x)$

{\bf Proof:} 
We utilize mathematical induction based on the index $n$.
The $n=1$ case is trivial, and the $n=3$ case can be proven by brute force.
Suppose that our claim holds for all positive odd integers less than or equal to some odd integer $n \ge 3$.
Then, we can apply Eq.~\eqref{eq:Recursion2} multiple times to obtain the recursion relation that involves only odd indices, i.e., $b_{n+2}$, $b_n$, and $b_{n-2}$:
\begin{widetext}
\begin{align}
b_{n+2} &= \sqrt{1-x^2} (e^{i \varphi_{n+1}} -1) b_{n+1} + e^{i \varphi_{n+1}} b_{n} 
\nonumber \\
&= \sqrt{1-x^2} (e^{-i \phi_{n+1}} -1) \left[ \sqrt{1-x^2} (e^{i \phi_{n}} -1) b_{n} + e^{i \phi_{n}} b_{n-1} \right] + e^{-i \phi_{n+1}} b_{n} 
\nonumber \\
&= e^{-i \phi_{n+1}} \left[1 - (1-e^{i \phi_{n+1}}) (1 - e^{i \phi_n}) (1-x^2) \right] b_n - (1-e^{-i \phi_{n+1}}) e^{i \phi_n} \sqrt{1-x^2} b_{n-1} 
\nonumber \\
&= e^{-i \phi_{n+1}} \left[1 - (1-e^{i \phi_{n+1}}) (1 - e^{i \phi_n}) (1-x^2) \right] b_n - e^{-i \phi_{n+1}} (1-e^{i \phi_{n+1}}) e^{i \phi_n} \frac{b_n - e^{-i \phi_{n-1}} b_{n-2}}{1 - e^{-i \phi_{n-1}}},
\end{align}
\end{widetext}
where we have used the induction hypothesis.
Let us now define $\tilde{b}_n=e^{i (\phi_2 + \phi_4 + \ldots \phi_{n-1})} b_{n}$, which satisfies the following recursion relation:
\begin{widetext}
\begin{align}
\tilde{b}_{n+2} &= \left[1 - (1-e^{i \phi_{n+1}}) (1 - e^{i \phi_n}) (1-x^2) \right] \tilde{b}_n  - (1-e^{i \phi_{n+1}}) e^{i \phi_n} \frac{\tilde{b}_n - \tilde{b}_{n-2}}{1 - e^{-i \phi_{n-1}}} 
\nonumber \\
&= \left[ (1- e^{i \phi_{n+1}} ) (1- e^{i \phi_n} ) x^2 + e^{i \phi_{n+1}} \right] \tilde{b}_n + (1-e^{i \phi_{n+1}}) e^{i \phi_n} \frac{\tilde{b}_n - e^{i \phi_{n-1}} \tilde{b}_{n-2}}{1 - e^{i \phi_{n-1}}} ,
\label{eq:Recursion3}
\end{align}
\end{widetext}

We can perform similar calculations to obtain the recursion relation involving $a_{n+2}$, $a_n$, and $a_{n-2}$ by using Eq.~\eqref{eq:Recursion1}:
\begin{widetext}
\begin{align}
a_{n+2} &= x (1- e^{i \phi_{n+1}} ) a_{n+1} + e^{i \phi_{n+1}} a_{n} \nonumber \\
&= x (1- e^{i \phi_{n+1}} ) \left[ x (1- e^{i \phi_{n}} ) a_{n} + e^{i \phi_{n}} a_{n-1} \right] + e^{i \phi_{n+1}} a_{n} 
\nonumber \\
&= \left[ (1- e^{i \phi_{n+1}} ) (1- e^{i \phi_n} ) x^2 + e^{i \phi_{n+1}} \right] a_n+ (1-e^{i \phi_{n+1}}) e^{i \phi_n} x a_{n-1} 
\nonumber \\
&= \left[ (1- e^{i \phi_{n+1}} ) (1- e^{i \phi_n} ) x^2 + e^{i \phi_{n+1}} \right] a_n + (1-e^{i \phi_{n+1}}) e^{i \phi_n} \frac{a_n - e^{i \phi_{n-1}} a_{n-2}}{1 - e^{i \phi_{n-1}}}.
\label{eq:Recursion4}
\end{align}
\end{widetext}
As one can see from the comparison between Eqs.~\eqref{eq:Recursion3} and \eqref{eq:Recursion4},  $a_n$ and $\tilde{b}_n$ satisfy the same recursion relation for odd indices.
Therefore, $a_n=\tilde{b}_n=e^{i (\phi_2 + \phi_4 + \ldots \phi_{n-1})} b_{n}$.
Noting that $\phi_n=-\varphi_n$ for even $n$, we can finally prove Eq.~\eqref{eq:Final_goal}.

\section{Optimization of the iteration length for amplitude amplification}
\label{appen:L-condition}

We begin with the following condition described in the main text:
\begin{align}
\sqrt{1-{\cal W}}/\gamma \le 1.
\label{eq:L-condition1}
\end{align}
Above, $\gamma$ is related to the desired tolerance, $\delta$, as follows: 
\begin{align}
T_L(1/\gamma)= 1/\delta ,
\label{eq:gamma_delta_relationship1}
\end{align}
where $T_L$ is the $L$-th Chebyshev polynomial of the first kind.

Equation~\eqref{eq:L-condition1} can be rewritten as follows:
\begin{align}
{\cal W} \ge 1-[T_{1/L}(1/\delta)]^{-2},
\label{eq:L-condition2}
\end{align}
where we have utilized the inverse of Eq.~\eqref{eq:gamma_delta_relationship1}:
\begin{align}
1/\gamma= T_{1/L}(1/\delta).
\label{eq:gamma_delta_relationship2}
\end{align}
Note that Eq.~\eqref{eq:gamma_delta_relationship2} can be obtained from the definition of the Chebyshev polynomial of the first kind:
\begin{align}
T_n(x)= 
\begin{cases}
\cos{(n\cos^{-1}{(x)})} & \text{if } |x|\leq 1 \\
\cosh{(n\cosh^{-1}{(x)})} & \text{if } |x| > 1 
\end{cases}
\label{eq:Chebyshev}
\end{align}

Now, assuming that $L \gg 1$ and $\delta \ll 1$, we can expand $T_{1/L}(1/\delta)$ as shown below:
\begin{align}
T_{1/L}(1/\delta) &= \cosh{\left(\frac{1}{L}\cosh^{-1}(1/\delta)\right)} 
\nonumber \\
&\approx 1+\frac{1}{2}\left( \frac{1}{L} \cosh^{-1}(1/\delta) \right)^2.
\label{eq:T_expand1}
\end{align}
Next, we can also expand $\cosh^{-1}(1/\delta)$ as follows:
\begin{align}
\cosh^{-1}(1/\delta) \approx \ln(2/\delta),
\label{eq:arccos_expand}
\end{align}
which can be proved by applying the hyperbolic cosine function to both sides of the equation. 
Then, Equation~\eqref{eq:T_expand1} can be rewritten as follows:
\begin{align}
T_{1/L}(1/\delta) \approx 1+\frac{1}{2}\left( \frac{\ln{(2/\delta)}}{L} \right)^2,
\label{eq:T_expand2}
\end{align}
which can, in turn, be plugged into Eq.~\eqref{eq:L-condition2} to generate the following condition:
\begin{align}
{\cal W} \ge \left( \frac{\ln{(2/\delta)}}{L} \right)^2.
\label{eq:L-condition3}
\end{align}
Finally, Equation~\eqref{eq:L-condition3} reduces to the following expression of the optimization condition for $L$:
\begin{equation}
L \ge \frac{\ln(2/\delta)}{\sqrt{{\cal W}}}.
\label{eq:L-condition_final}
\end{equation}

\bibliographystyle{apsrev4-2}
\bibliography{references}

\end{document}